\documentclass[11pt]{revtex4-1}

\usepackage{graphicx}
\usepackage{bm}
\usepackage{amssymb}
\usepackage{color}

\begin{document}

\title{
A null test of General Relativity: New limits on Local Position Invariance and the variation of QCD scaled quark mass
}

\author{Neil Ashby}
\affiliation{National Institute of Standards and Technology\footnote{This work is a contribution of NIST and is not subject to U.S. copyright.}\\
325 Broadway, Boulder, CO 80305-3337, USA\\}
\affiliation{Department of Physics, University of Colorado\\
2000 Colorado Blvd., Boulder, CO 80309-0390, USA}

\author{Thomas E. Parker}
\affiliation{National Institute of Standards and Technology\\
325 Broadway, Boulder, CO 80305-3337, USA}

\author{Bijunath R. Patla\footnote{To whom correspondence should  be addressed:bijunath.patla@nist.gov}}
\affiliation{National Institute of Standards and Technology\\
325 Broadway, Boulder, CO 80305-3337, USA}

\begin{abstract}
We compare the long-term fractional frequency variation of four hydrogen masers that are part of an ensemble of clocks comprising the National Institute of Standards and Technology\,(NIST), Boulder, timescale with the fractional frequencies of primary frequency standards operated by leading metrology laboratories in the United States, France, Germany, Italy and the United Kingdom for a period extending more than 14 years. 
The measure of the assumed variation of non-gravitational interaction\,(LPI parameter, $\beta$)---within the atoms of H and Cs---over time as the earth orbits the sun, has been constrained to $\beta=(2.2 \pm 2.5)\times 10^{-7}$,  a factor of two improvement over previous estimates. Using our results together with the previous best estimates of $\beta$ based on Rb vs. Cs, and Rb vs. H comparisons, we impose the most stringent limit to date on the variation of scaled quark mass with strong(QCD) interaction to the variation in the local gravitational potential.
 For any metric theory of gravity
$\beta=0$.
\end{abstract}

\maketitle

General Relativity (GR) is one of the most successful theories of physics, explaining satisfactorily numerous phenomena of gravitation as well as many phenomena that would be inexplicable in a Newtonian universe, such as perihelion precession of the inner planets or gravitational frequency shifts.  We would have limited understanding of our universe---for example, the recession of distant galaxies or the early history and the subsequent evolution of the universe---without the help of GR\cite{einstein96,weinberg72,demille17}.  


In GR, 
space is not necessarily Euclidean nor does it necessarily stretch infinitely in all three directions.  Clock rates and measuring rod lengths may be affected by the amount of energy and momentum in the neighbourhood.  Alternative theories of gravity go even further, for example allowing clock rates to depend on the internal structure of the atoms with which the clocks are constructed, or allowing the results of similar experiments to differ if performed at remotely located places or times.  It is the rates of clocks on earth that we study in this paper, as the earth orbits the sun---over a period of more than 14 years.

Several far-reaching principles are embedded in Einstein's GR.  The general consensus is that any metric theory such as GR satisfies the Einstein Equivalence principle\,(EEP) that encapsulates three main principles\cite{Will14}: a)\,Local Position Invariance (LPI), b)\,Weak Equivalence Principle (WEP), and c)\,Local Lorentz Invariance (LLI). 

LLI states that the laws of physics must be independent of the velocity of the reference frame in which the laws are expressed; in other words, the laws of physics must be form-invariant with respect to transformations between relatively moving reference frames.  WEP requires that in a gravitational field, all objects---regardless of their internal composition---fall with the same acceleration. This principle is also known as the Universality of Free Fall (UFF); as a consequence, the results of experiments in a small laboratory having an acceleration $\vec{ a}$ must be the same as the results of similar experiments performed in a small laboratory in a gravitational field of strength $\vec{g}=-\vec{a}$.  LPI states that the outcome of an experiment must be independent of the position and orientation of the reference frame in which the experiment is performed.  LPI is the topic of the present study; the remainder of this paper assumes that both WEP and LLI are valid.

In an accelerated laboratory, if two otherwise identical clocks separated by height $h$ exchange photons, the photon frequencies will suffer first-order Doppler shifts due to the velocity difference that builds up during the propagation delay between clocks, because the speed of light is finite. This implies clocks at different gravitational potentials  will suffer frequency shifts that do not depend on the structure of the clocks. A comparison of the frequencies of two similar clocks at different locations, can be considered as a nonlocal gravitational experiment and understood within the framework of EEP.  
The gravitational redshift described above has been measured accurately to 120 parts per million\cite{vessot80}.

Local position invariance (LPI) assumes that the outcome of any local experiment that measures a nongravitational effect is independent of the spacetime location at which the experiment is performed. In our study, the hyperfine splitting in hydrogen and cesium atoms, arising from magnetic interactions between angular momenta, are the nongravitational interactions of interest.   
We look for variations in atomic transition frequencies arising from such interactions, as the earth orbits the sun, thereby changing the gravitational potential in which the transitions occur.  Thus if two clocks of different internal structures move together through a gravitational potential, their frequency ratio must be constant, otherwise their frequency shifts relative to a reference at a different gravitational potential would not be unique.

The advances reported here in testing LPI are complementary to at least one planned space based experiment for testing the postulates of metric theories of gravity.  Atomic Clock Ensemble in Space\,(ACES) comprising a H-maser and a Cs tube onboard the International Space Station\,(ISS) for comparing ground based clocks using microwave links with a tentative launch in 2018, will aim to test the gravitational redshift and LLI\cite{hess11}. The now called off, but nevertheless highly rated science experiment, Space-Time Explorer and QUantum Equivalence Principle Space Test\,(STE-QUEST) had plans to test the WEP using atom interferometry\cite{altschul15}. As clocks become more portable and space-qualified, one could foresee more such experiments planned well into the future.

The National Institute of Standards and Technology\,(NIST), in Boulder, Colorado, hosts five hydrogen masers and four commercial cesium standards as the basis of the timescale that provides---along with the U.S. Naval Observatory---official civil time UTC\,(NIST) for the United States.  By international convention, the exact frequency 9,192,631,770 s$^{-1}$ corresponding to the  hyperfine splitting in the ground state, $|F=3;m_F=0\rangle \leftrightarrow  |F=4;m_F=0\rangle$, of Cs$^{133}$ atoms held at a temperature of 0\,K provides the international system (SI) definition of the second. The definition is realized at major national laboratories through primary Cs-fountain frequency standards such as NIST-F1 or NIST-F2, which are run intermittently and are used to improve the long-term stability of the NIST  timescale,  and to help calibrate International Atomic Time\,(TAI).

The Cs-fountains and the H-masers are all considered to be ``local," with spatial separations that are relatively small, in the locally inertial frame centered on the earth as it falls freely in the solar system's gravitational field, see Figure\,1.
\begin{figure}[!htb]  
\begin{center}
\includegraphics[width=0.8\textwidth]{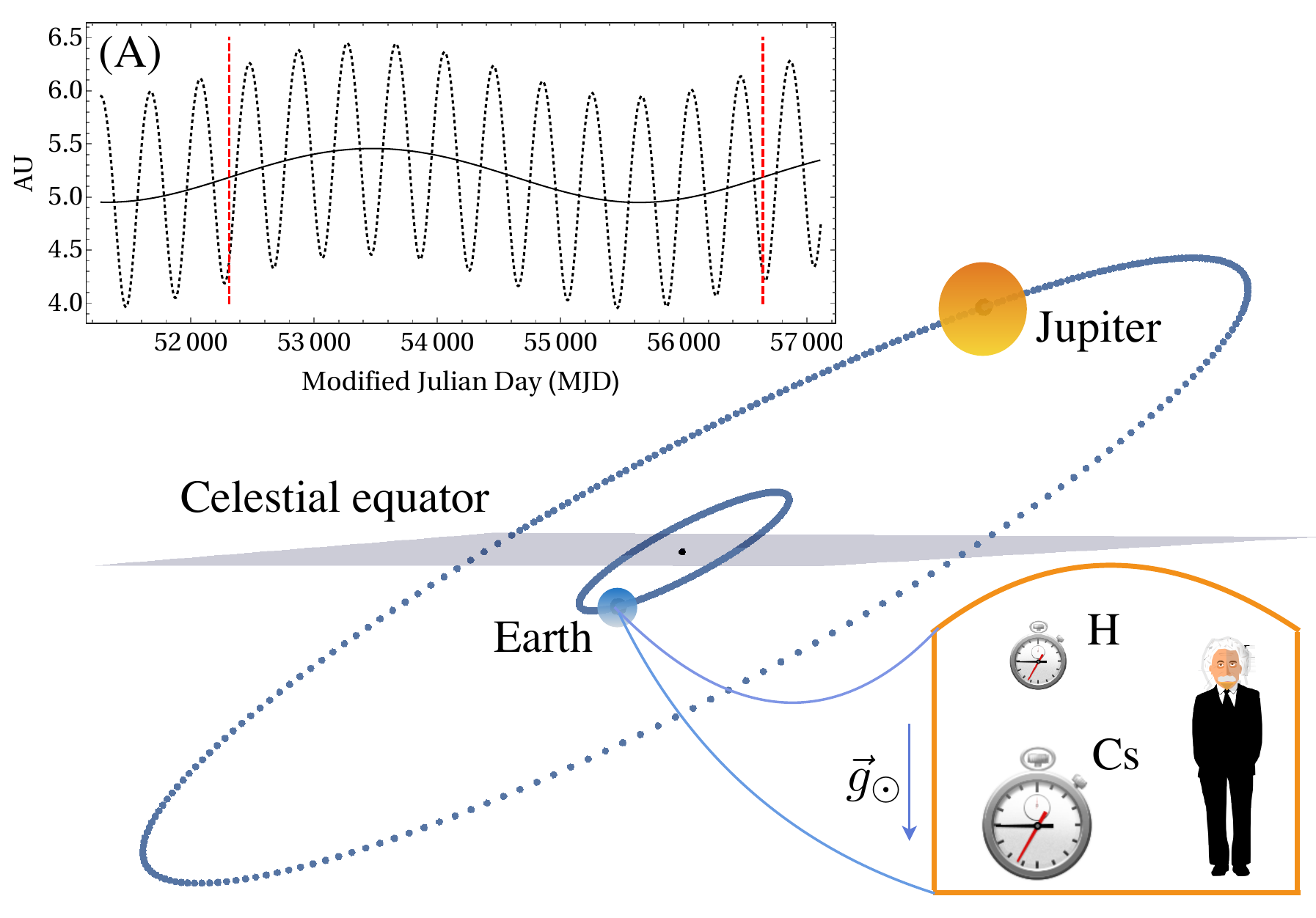}
\caption{ Position of earth and Jupiter on MJD56263 (December 02, 2012).  Planets orbiting the sun are not drawn to scale.  The earth is in freefall in the solar system gravitational field with an acceleration $g_\odot\sim0.006$\,m/s$^2$. The earth, including the H-masers and Cs-fountains may be conceived as an Einstein elevator (negligible tidal forces). 
Inset\,(A): Distance (in AU) of the barycenter of Jupiter from the sun's center  and the distance between the barycenters of Jupiter and  earth are represented by solid and dotted curves.  The time period between the dashed vertical lines is $\sim 11.86$ years, the orbital period of Jupiter. 1\, AU is roughly the mean distance between the earth and sun, which now has a fixed assigned value, see Table 1.
Jupiter's radius is $\sim$\,11 times that of earth's radius and the orbital radius  is $\sim$\,5.2 times larger than the earth's orbital radius.}  
\end{center}
\end{figure} 
This is in compliance with relativity principles that require LPI to be true only ``locally."
In order to test relativity to higher precision, accurate measurements of time and frequency---with regard to clock comparisons---are critical.  In the past decades, technological advancements in precision metrology have made possible time and frequency measurement with higher precision and better stability\cite{sullivan01}.                 

H-masers and Cs-fountains are atomic clocks that exploit small differences in the energy levels of the internal states of the atoms---making them very accurate and stable frequency standards. Two different species of atoms, H and Cs in this study, have different internal structures, in terms of the neutron to proton ratio ($N/Z$), and in the electromagnetic contribution to the binding energy ($\propto Z^2$) for each atomic species\cite{dicke64}. $N$ and $Z$ for hydrogen are 0 and 1. For cesium, $N$ and $Z$ are 78 and 55.  

Once the relative frequency offset and frequency drift of two clocks are corrected for, LPI requires that clock comparisons (frequency ratios) should remain the same as the clocks move together arbitrarily through a gravitational potential.  Such a comparison does not involve direct time transfer between space-borne clocks and clocks on the ground, nor does it require the clocks to be accurate in frequency. For such tests, the longer-term stability (stability for an orbital period or longer) of the clocks is relevant, and it is clear that the same control of systematic effects that yields high accuracy also leads to high stability.  Orbiting clocks have varying  position and velocity states in a gravitational field.  Local position invariance can be tested by studying variations in the frequency difference as the orbit radius and orbital speed vary. 


A change in the gravitational potential at the location of a clock, according to various alternatives to GR, causes a fractional frequency shift in the clock
\begin{equation}
\Delta f/f=(1+\beta)\Delta \Phi/c^2,
\label{eq_fracqb}
\end{equation}
where $\Delta \Phi$ is the change in gravitational potential,  $\Delta f$ is the change in frequency $f$ of the clock, and $c$ is the speed of light in vacuum. The parameter $\beta$ measures the degree of violation of LPI; in GR $\beta=0$.  The eccentricity of the earth's orbit ($e=0.0167$) provides sufficient variation of the distance separating the earth and the sun to assess a possible correlation between the annual variation of gravitational potential and the corresponding frequency offset introduced in the clocks.   The size of the earth is extremely small compared to the variation of earth-sun distance, so the gravitational redshifts arising from the solar potential differences between the two clocks 
positioned at different locations on the earth are very nearly the same, and in GR are canceled by relativistic effects arising from free fall of the clocks.  The non-gravitational contribution to the difference of the fractional frequency shifts  of two different clock types, H and Cs, is
\begin{equation}
\Delta f/f|_{\rm H}-\Delta f/f|_{\rm Cs}\equiv\Delta f/f|_{\rm H-Cs}=\beta\Delta \Phi/c^2,
\label{eq_fracq2}
\end{equation}
where $\beta=(\beta_{\rm H}-\beta_{\rm Cs})$.
In a null test of a metric theory of gravity such as GR, a measurement would put an upper limit on the absolute value of $\beta$.   

While the present work builds upon and extends the work of Ashby et~al.\cite{ashby07}, similar experimental tests of LPI have been a topic of interest for a very long time.  In 1978 Turneaure et~al. compared two hydrogen masers with a set of three superconducting cavity-stabilized oscillators as the solar gravitational potential changed due to earth rotation\cite{turneaure83}.  Measurements over a ten-day period were consistent with LPI and EEP at about the two percent level.  Godone et~al.\cite{godone95} compared Mg and Cs standards for 430 days and were able to improve on Turneaure's result by a factor of almost 20.  


In 2012 Guena and co-workers at SYRTE were able to compare Cs and Rb laser-cooled atomic fountain clocks over a period of 14 years, using variations in the solar gravitational potential to place significant limits on the rate of change of the fractional frequency difference of the two clocks, to obtain $\beta|_{\rm Rb-Cs}=(1.1\pm 10.4)\times 10^{-7}$\cite{guena_dada_12}. In 2013 Peil and co-workers at the Unites States Naval Observatory used continually running clocks (Rb fountains and H-masers) for 1.5 years and reported a value of $\beta|_{\rm Rb-H}=(-2.7\pm 4.9)\times 10^{-7}$\cite{peil13}.
For comparison with this we quote the result of previous comparisons at NIST for H and Cs $\beta|_{\rm H-Cs}=(1.0\pm 14) \times 10^{-7}$\cite{ashby07}.

For the present study, we chose eight Cs fountain primary frequency standards: IEN-CsF1 from Istituto Nazionale di Ricerca Metrologica\,(INRIM), Torino, Italy\cite{torinof1}; NIST-F1 from National Institute of Standards and Technology\,(NIST), Boulder, USA\cite{nistf1}; PTB-CsF1 and -CsF2 from Physikalisch-Technische Bundesanstalt\,(PTB), Braunschweig, Germany\cite{ptbf1,ptbf2};
NPL-CsF1 and -CsF2 from National Physical Laboratory, Teddington, UK\cite{nplf}; SYRTE-CsFO1 and -CsFO2 from
 Syst{\`e}mes de R{\'e}f{\'e}rence Temps-Espace (SYRTE), Paris, France\cite{syrtef}.
 The fractional frequency shifts of Cs primary frequency standards, referenced to the geoid, are reported to the Bureau International des Poids et Mesures, S{\`e}vres, France, ordinarily after each evaluation and are available from the BIPM ``Circular T"\cite{bipm}.
 
 The four NIST hydrogen masers that are used in this study are labeled S2 through S5. The masers are housed within environmentally controlled chambers and monitored for fluctuations in pressure, temperature, magnetic field, and humidity. The frequency shifts introduced by the environmental variables are computed based on measured frequency sensitivities  corresponding to each variable, for each maser. In general, the corrections for environmentally-caused frequency shifts are of the order of $10^{-16}$ through $10^{-13}$. For these masers, the temperature corrections were the most consequential and during some epochs were as high as $10^{-13}$.

Environmental factors affecting H-masers are studied in detail in Parker(1999)\cite{parker99} and the impact of frequency transfer noise in comparing masers and Cs-fountains are described in Parker et~al.(2005)\cite{parker05}, also see \cite{ashby07}.  The difference of frequency shifts of fountains versus a typical maser with time, after correcting for changes in environmental variables, are plotted in Figure\,2 with MJD (Modified Julian Date) on the abcissa. 
\begin{figure}[ht]  
\begin{center}
\includegraphics[width=0.95\textwidth]{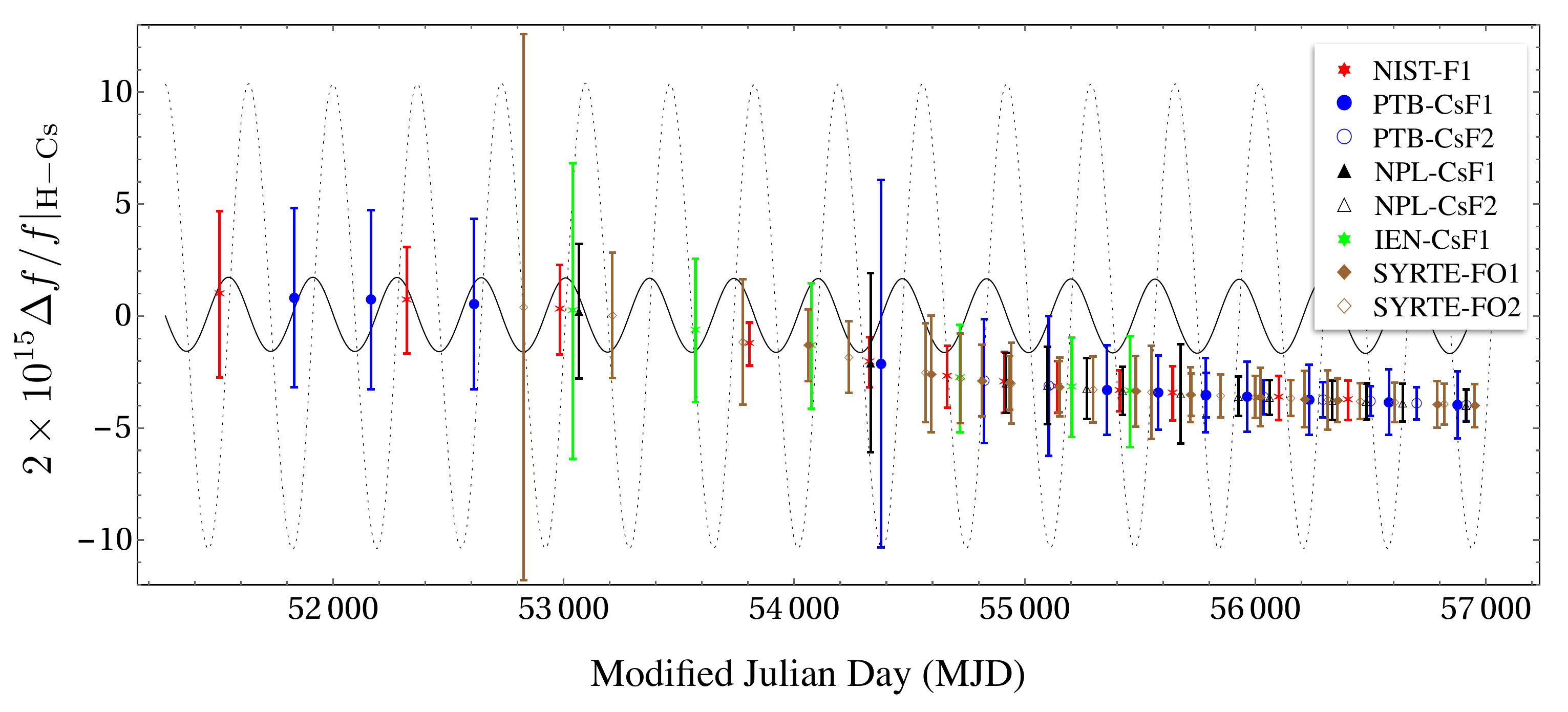}
\caption{ Frequency shifts of H and Cs for maser S3 for MJD 51508 (November 11, 1999) through MJD 56959 (October 29, 2014): The solid curve is the change in gravitational potential, $\Delta \Phi/c^2$, due to the sun and Jupiter as a function of time and the dotted curve is the total rate of change of gravitational potential,  $\Delta \Phi/dt/c^2[{\rm yr}^{-1}]$. Gravitational potential variations are scaled by a factor of $10^{10}$, instead of the factor $2\times10^{15}$\,(see y-axis) used for scaling the frequency shifts. The data are the differences in  fractional frequency shifts of H and Cs after accounting for environmental corrections for maser S3. There is clear evidence of component aging related drift for this maser and this behavior is typical for all four masers considered. The fractional frequency differences are suppressed by a factor of 100 to be able to show the estimated uncertainties and the fractional frequency differences in the same plot.
Only 20\% of the data are shown to avoid blotting out the curves, for more information, see Table~3 of appendix.   One of the reasons this study is an improvement over the previous effort is there is significantly more fountain data after MJD54000\cite{ashby07}. The identifiers used for some of the Cs-fountains or the host laboratories may have changed over the years.}  
\end{center}
\end{figure}  
The stated uncertainty is a combined measure of the Cs-fountain uncertainty and the uncertainty in   the frequency transfer from the Cs-fountain to the maser. Since all the masers are housed in the same location, the stated uncertainty is the same for all the masers; the corrections on frequency fluctuations due to changes in environmental factors are different.
In addition to environmental factors, the masers experience long-term drifts that are related to component aging\cite{lewis91,parker99}. Frequency shifts for all H-masers are referenced to the location of NIST, Boulder.

A third order polynomial was used to fit the fractional frequency difference data for phenomenologically accounting for component aging and any other systematic shifts that can be as high as $\Delta f/f \approx 10^{-14}$~/year. This should not affect the correlation sought between the residuals of the fit and sinusoidal solar potential variation. Roughly the data are split into three segments of 5 years each such that the residuals from the fit conform to a normal distribution. Data segmentation was necessitated by the requirement to keep the number of free parameters in the fit to a minimum and to keep the fitting functions simple.  A plot of the residuals is shown in Figure\,3; only 20\,\% of the data points are shown to avoid blotting out the curves.
\begin{figure}[ht]  
\begin{center}
\includegraphics[width=0.95\textwidth]{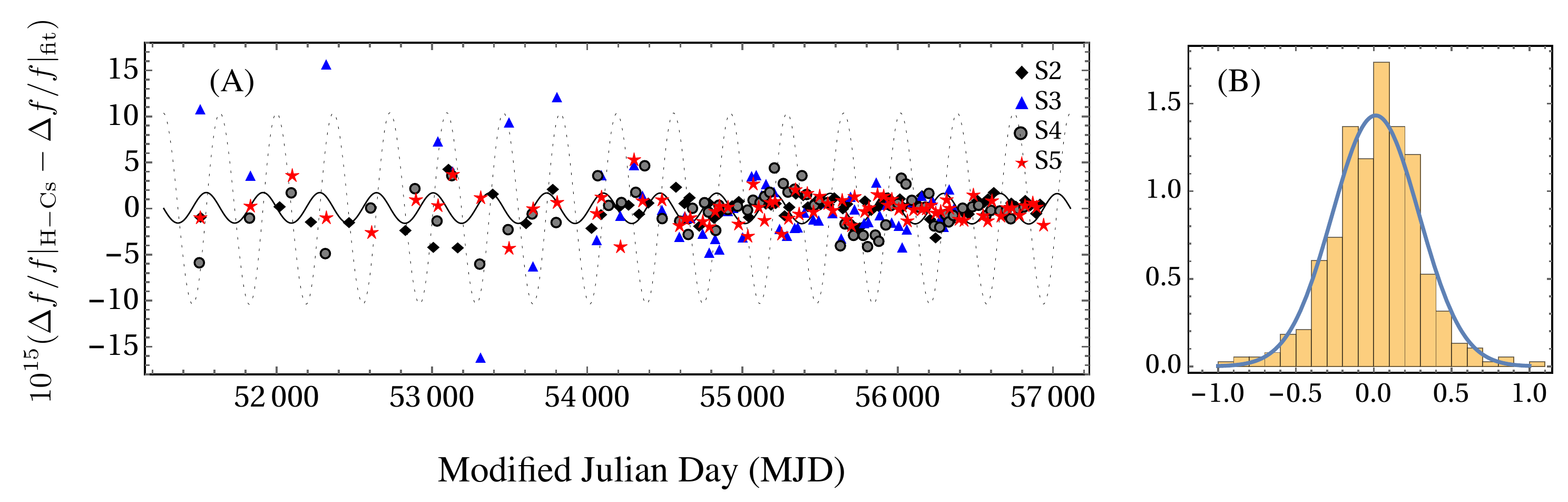}
\caption{Panel\,(A): Residuals after fitting the fractional frequency difference data with third order polynomials: Solid and dotted curves are the gravitational potential variation and its time derivative, respectively. S3 has a larger scatter compared to other masers. As in Figure\,2, the solid and dotted curves are the gravitational potential variations.
Panel\,(B): Histogram of the 380 residuals for maser S2---as an example---scaled to a normal probability distribution function (vertical axis). 
The horizontal axis is the residuals from the third order polynomial fit
scaled to unity so that the probability distribution function(solid curve) and the data can be presented in the same plot. The actual values
range between $\approx\pm 5.0\times 10^{-15}$.}
\end{center}
\end{figure} 

The spatial variation of the gravitational potential is computed using DE430 planetary and lunar ephemerides from JPL\cite{folkner14}. We calculate the gravitational potential of Jupiter by computing the distance between the Jupiter's barycenter  and the earth-moon barycenter. Jupiter's mass is about a thousand times smaller than that of the sun but twice that of the rest of  the planets in the solar system. The precision of DE430 ephemerides, which is  sub-kilometer within the inner solar system, allows us to realistically account for the effect of Jupiter's gravitational field. The effect of Jupiter's potential wasn't considered in previous studies reporting tests of LPI.

Each data point in Figure\,2 is the average of 10 to 40 days corresponding to the time when the Cs-fountains were evaluated. The time stamp of a data point is the midpoint of the evaluation period. The value of the gravitational potential assigned to this midpoint is the average over the fountain evaluation period. The solar potential is evaluated using the distance between the sun's center and the earth-moon barycenter. Jupiter's potential is evaluated using the distance between the earth-moon barycenter  and the barycenter of Jupiter, also see Figure 1.
The values of all constants used in performing the above calculation and assumed to be non-varying are given in Table~1.
\begin{table}[ht]
\centering
\caption{Constants and their values}
\label{tab1}
\begin{tabular}{ll}
\hline\hline
constants  &  value\\ \hline
\\[-2.0ex]
astronomical unit, AU & 149597870.691 km\\
speed of light in vacuum, $c$  & 299792458 m\,s$^{-1}$ \\
standard gravitational parameter of sun, $GM_S$ & $1.327124420\times 10^{20}\;$m$^3$s$^{-2}$ \\
Newton's gravitational constant, $G$ & $6.67428\times 10^{-11}\;$m$^3$kg$^{-1}$s$^{-2}$\\
ratio of mass of sun and mass of Jupiter, $M_S/M_J$ \hspace{4cm} & 1047.348644\\ \hline\hline
\end{tabular}
\end{table}
For each H-maser, the amplitude of the LPI parameter  $\beta$ is computed by using the residuals from the polynomial fits and the combined potential variation due to the sun and Jupiter for the epoch corresponding to the fountain evaluation. We look for correlation in the residuals with a fixed phase and period corresponding to the variation in the total potential, using Eq.~(\ref{eq_fracq2}).  
The uncertainty is obtained by performing a standard least squares fit\,(all Cs-fountains are assigned equal weights) of the data. The results for the amplitude and uncertainty for all the four masers are combined to obtain the final result (see appendix for more details on data analysis)
\begin{equation}
\beta|_{\rm H-Cs}=(2.24\pm 2.48)\times 10^{-7}.
\label{delbeta}
\end{equation}
This study improves the uncertainty in $\beta$ by more than a factor of five compared to our previous study in 2007, and imposes a stricter contraint on the uncertainty in $\beta$ reported for  any two pairs of atoms\cite{ashby07,guena_dada_12,peil13}. The inclusion of Jupiter's gravitational potential had an effect only on the third significant digit with a contribution of the order of a percent. For the entire data set, 
the uncertainty in the estimation of the gravitational potential variation using the planetary ephemerides is three orders of magnitude smaller than the combined stated uncertainty due to frequency transfer. Therefore, in our calculations leading to Eq.~(\ref{delbeta}), we have neglected the uncertainty in the estimation of the gravitational potential.

The variation of nongravitational interactions with time implies that in the low energy universe certain fundamental constants could change if LPI is violated. 
It is this aspect of LPI that takes one of the postulates applicable to metric theories of gravitation closer to general relativity through an additional requirement: the principle of general covariance. It states that the laws of physics ought to be expressible in a coordinate independent formalism; constants of nature comprise one part of that story.
In the following paragraphs, we'll use the result of Eq.~(\ref{delbeta}) together with the previous best estimates of $\beta$ to place the best constraints to date on the variations of two fundamental constants. 

In the realm of atomic physics, matter and its interaction with fields may be parameterized  in terms of the masses of quarks, mass of electron, fine structure constant $\alpha$, and quantum chromodynamics (QCD) energy scale parameter $\Lambda_{\rm QCD}$ at which the QCD coupling begins to diverge\cite{uzan03}. $\alpha$ describes electromagnetic interactions in matter and $\Lambda_{\rm QCD}$ measures the strong interaction.
For example, $\alpha \sim 1/137$, can be interpreted as the ratio of the speed of an electron in the Bohr atom to the speed of light (photon is the force carrier for the electromagnetic force) in vacuum.
Dicke, in his 1963 lectures, conjectured that $c^2 (d\alpha/dt) \approx \alpha^2 (d\Phi/dt)$, where $d\Phi/dt$ is the time variation of the gravitational potential \cite{dicke64}.


The difference in frequency shifts due to hyperfine splitting for a pair of clocks may be recast as a variation of the ratio of the frequencies, which is related to the variation of the fundamental constants by the formula\cite{fischer_alpha_04, flambaum06,dinh_quark_09,guena_dada_12}
\begin{equation}
d\ln \left( f_{\rm A}/f_{\rm B}\right)=\Delta K_{\alpha}\,d\ln \alpha+ \Delta K_q\,d\ln X_q,
\label{eq_totvar}
\end{equation}
where $\alpha$ is the fine structure constant,   $X_q=m_q/\Lambda_{\rm QCD}$ is the ratio of the light quark mass to the QCD scale. $K_{\alpha}$ and $ K_q$ are the relative sensitivities of the hyperfine relativistic factor and nuclear magnetic moment to the variation  of  $\alpha$ and $X_q$ respectively. 
Since the hydrogen masers used in this study are susceptible to drifts whose origins are not well understood, over periods that are of the order of few years, 
below we present a formalism to constrain  $d\ln \left( f_{\rm H}/f_{\rm Cs}\right)$. 

The ratio of hyperfine frequencies of two atomic species are related to the spatial variation of gravitational potential, from Eq.~(\ref{eq_fracq2}), which can also be written as: 
\begin{equation}
d\ln \left( f_{\rm H}/f_{\rm Cs}\right)=(\beta \Delta \Phi)/c^2,
\label{constvar1}
\end{equation}
where $c$,  $\beta$ and $\Delta \Phi$ are the same quantities as in Eq.~(\ref{eq_fracqb}) and (\ref{eq_fracq2}). 
In order to constrain $\alpha$ and $X_q$ individually, first we note that 
\begin{equation}
\delta \alpha/\alpha=k_\alpha\delta(\Phi/c^2)\quad {\rm and} \quad \delta X_q/X_q=k_q\delta(\Phi/c^2)
\label{eq_coupling}
\end{equation}
where $k_\alpha$ and $k_q$ are dimensionless coupling constants linking the variation of $\alpha$ and $X_q$ to the variation of the gravitational potential. Using Eq.~(\ref{eq_coupling}) in Eq.~(\ref{eq_totvar}) and rearranging the terms, we obtain equations of the form
\begin{equation}
\beta|_{\rm H-Cs}=\Delta K_\alpha k_\alpha + \Delta K_q k_q.
\label{eq_ks}
\end{equation}
We'll use the previous best estimates for $\beta$ involving clock transitions that depend on hyperfine splitting analyzed in this study for solving for the dimensionless coupling constants, see Table~\ref{tab2}.
\begin{table}[ht]
\centering
\caption{Comparing previous best estimates on $\beta$  with this study. The values of the differences in $K_\alpha$ and $K_q$ are from Flambaum and Tedesco, 2006\cite{flambaum06}.}
\label{tab2}
\begin{tabular}{lllllr}
\hline\hline
\# & Reference  & $\beta_A-\beta_B$ & $A,B$ & $\Delta K_{\alpha}$ & $\Delta K_q $ \\ 
\hline
\\[-3.0ex]
(i) \hspace{1cm} & Gu\' ena et al., 2012\cite{guena_dada_12}\hspace{1cm} & $(0.11\pm 1.04)\times 10^{-6}$\hspace{1cm} & Rb,Cs\hspace{1cm} & $-0.49$\hspace{1cm} & $-0.025$   \\
(ii) & Peil et al., 2013\cite{peil13} & $(-2.7\pm 4.9)\times 10^{-7}$ & Rb,H & $\;\;\;0.34$ & $\;\;\;0.084$   \\
(iii) & this work & $(2.24\pm 2.48)\times 10^{-7}$ & H,Cs & $-0.83$ & $-0.110$   \\
\hline\hline
\end{tabular}
\end{table}
Using the entries of Table~\ref{tab2} in Eq.~(\ref{eq_ks}), yields two independent sets of values for $k_\alpha$ and $k_q$ for equations involving pairs (i) and (iii), and (ii) and (iii) of Table~\ref{tab2}. The equally weighted averages of the two values for both $k_\alpha$ and $k_q$ yields:
\begin{equation}
k_\alpha=(0.70 \pm 1.8)\times 10^{-7} \quad {\rm and } \quad
k_q=(-25 \pm 21)\times 10^{-7}. 
\label{eq_kfinal}
\end{equation}
The previous best estimates for $k_q=(3.8\pm 4.9)\times10^{-6}$ was reported by Peil at al.(2013)\cite{peil13}. More recently, Dzuba and Flambaum (2017) report a slightly better value of $k_\alpha=(-0.53\pm 1.0)\times10^{-7}$\,\cite{flambaum17}.
Our results are an improvement over the previous estimates of $k_q$ by a factor of two.
The combined annual variation of  gravitational potential due to sun and Jupiter based on the ephemerides is $3.313\times 10^{-10}$. Using this value in Eq.~(\ref{eq_coupling})
\begin{equation}
\dot{\alpha}/\alpha=(2.3 \pm 6.0)\times 10^{-17}/{\rm yr} \quad {\rm and } \quad
\dot{X_q}/X_q=(-8.3\pm 7.0)\times 10^{-16}/{\rm yr}.
\label{eq_xfinal}
\end{equation}
Godun et al.(2014)\cite{godun14} estimated  $\dot{\alpha}/\alpha=(-0.7 \pm 2.1)\times 10^{-17}/{\rm yr}$ from direct measurements---a factor of three better than the result presented here. Gu\'{e}na et al. (2012)\cite{guena_dada_12} had set the previous best estimates for $\dot{X_q}/X_q=(0.14\pm 9.0)\times 10^{-16}/{\rm yr}$, as correctly inferred by Huntemann et al. (2014)\cite{hunt14}.

Since LPI---as a postulate of GR---is more general than any experiment involving only two atomic species, combining the values of LPI parameters from Table~\ref{tab2}, we obtain the weighted average
\begin{equation}
\beta=(2.2 \pm 2.2)\times 10^{-7},
\label{eq_lpi_final}
\end{equation}
with the assigned weights that are equal to the inverse of the square of the uncertainties.



A null hypothesis ($\beta=0$) is a necessary condition for any metric theory like general relativity to be valid; since all experiments have finite errors, no experiment can serve as a sufficient condition\cite{dicke64}.
By deriving new limits on the variations of two fundamental constants, we were able to extend the applicability of the null hypothesis of LPI for validating metric theories, that are a more general class of theories, to GR.
The implications of varying fundamental constants in the context of unified theories and alternatives to GR are detailed in Uzan(2003)\cite{uzan03}. 


We note that using three masers instead of four made only a small difference in the estimation of $\beta$. More data is unlikely to yield stricter constraints. Owing to the long-term drifts that are typical in H-masers, there is not much likelihood for improving the uncertainty in the LPI parameter using H-masers and Cs-Fountains.
Future improvements  are most likely to come from comparisons of optical clocks, which might perform at least two orders of magnitude better---only limited by the uncertainty in the estimation of the total gravitational potential variation---than  comparisons between H-masers and Cs-fountain standards as the noise contributions in optical clocks are better understood as the performance of these clocks continue to improve\cite{ludlow15}. 

Of the many challenges in comparing different optical clock types, up until recently, the main ones are the availability of frequency links with stability and frequency transfer uncertainty comparable to the best optical clocks, and availability of robust clocks capable of running  simultaneously over periods that match the earth's orbital period. 
An example of the improvement in the development of fiber links is the recently commissioned 1415~km telecom fiber link connecting Paris and Braunschweig\cite{lisdat16}. Work is also underway to compare the NIST ytterbium clock and JILA strontium clock using a fiber link\cite{bloom14,nicholson15,schioppo17}. These optical clocks and fiber links are two important aspects of any future experiments that are certain to improve the results presented in this paper, at which time the effects of gravitational perturbation from Jupiter won't be negligible as it was for this study\cite{tobar2013}. 
  
\section{appendix}
\noindent A more detailed procedure for obtaining $\beta$ is described below. For each maser, optimizing Eq.~(\ref{eq_fracq2}) 
\begin{equation}
\sum_i \frac{d}{d\beta}\left(\Delta_i-\beta\Delta\Phi_i/c^2\right)^2=0,
\end{equation}
where $\Delta_i=\Delta f/f|_{\rm H-Cs}- \Delta f/f|_{\rm fit}$,
yields
\begin{equation}
\beta_j=c^2\frac{\sum_{_i} \Delta_i \Delta \Phi_i }{\sum_{_i} \Delta \Phi_i^2},
\end{equation}
where the index $i$ is the time stamp label for a data point. For example, $i$ varies from $1$ to $380$ for maser S2 ($j=1$), see Table~\ref{tab3}. $\Delta \Phi$ is the change in total gravitational potential. The maximum and minimum values of frequency difference for each maser vs. fountain after correcting for environmental effects are also given in Table~\ref{tab3}.
Maser frequency drifts are analyzed and quantified in Ashby et al.(2007)\cite{ashby07}.
\begin{table}[ht]
\centering
\caption{Comparing values of $\beta$ from masers S2-S5}
\label{tab3}
\begin{tabular}{lrcllcc}
\hline\hline
maser & \# data  & MJD$_{\rm max}$ & $\Delta f/f|_{\rm max}$ & MJD$_{\rm min}$ & $\Delta f/f|_{\rm min}$ & $\beta_j$ \\ 
\hline
\\[-3.0ex]
S2&\hspace{1cm}  $380$& \hspace{0.5cm}$56959$\hspace{0.3cm} & $4.001\times 10^{-13}$&\hspace{0.3cm} $53164$ &\hspace{0.5cm} $2.623\times 10^{-15}$ &\hspace{0.3cm} $(1.55\pm 4.74) \times 10^{-7}$
\\
S3&\hspace{1cm}  $314$& \hspace{0.5cm}$56394$\hspace{0.3cm} & $2.226\times 10^{-12}$&\hspace{0.3cm} $53931$ &\hspace{0.5cm} $7.975\times 10^{-13}$ &\hspace{0.3cm} $(7.24\pm 5.57) \times 10^{-7}$
\\
S4&\hspace{1cm}  $363$& \hspace{0.5cm}$56824$\hspace{0.3cm} & $6.476\times 10^{-13}$&\hspace{0.3cm} $51508$ &\hspace{0.5cm} $1.364\times 10^{-13}$ &\hspace{0.3cm} $(3.47\pm 4.94) \times 10^{-7}$
\\
S5&\hspace{1cm}  $384$& \hspace{0.5cm}$54376$\hspace{0.3cm} & $2.208\times 10^{-13}$&\hspace{0.3cm} $51508$ &\hspace{0.5cm} $7.387\times 10^{-14}$ &\hspace{0.3cm} $(-1.81\pm 4.75) \times 10^{-7}$
\\
\hline\hline
\end{tabular}
\end{table}
The uncertainty for an individual maser correlated with the gravitational potential is 
\begin{equation}
\langle \delta\beta_j^2 \rangle=c^4\frac{\langle\sum_{_{ik}} \delta(\Delta_i) \Delta \Phi_i \delta(\Delta_k) \Delta \Phi_i\rangle}{\left(\sum_{_{i}} \Delta \Phi_i^2\right)^2},
\end{equation}
where $\langle\delta(\Delta_i)\delta(\Delta_k)\rangle=\delta_{ik}\delta(\Delta_i)^2$, to give uncorrelated
\begin{equation}
\delta\beta_j=c^2\frac{\left(\sum_{_i}\delta(\Delta_i)^2\Delta\Phi_i^2\right)^{1/2}}{\sum_{_{i}} \Delta {\Phi_i}^2}
\end{equation}
where $\delta(\Delta_i)$ is the combined uncertainty (Cs-fountain and frequency transfer from fountain to maser). There has been significant improvement in the reported uncertainties in the last 7 years of the data set compared to the first 7 years (before MJD 54000), see Fig.~4.
\begin{figure}[ht]  
\begin{center}
\includegraphics[width=0.6\textwidth]{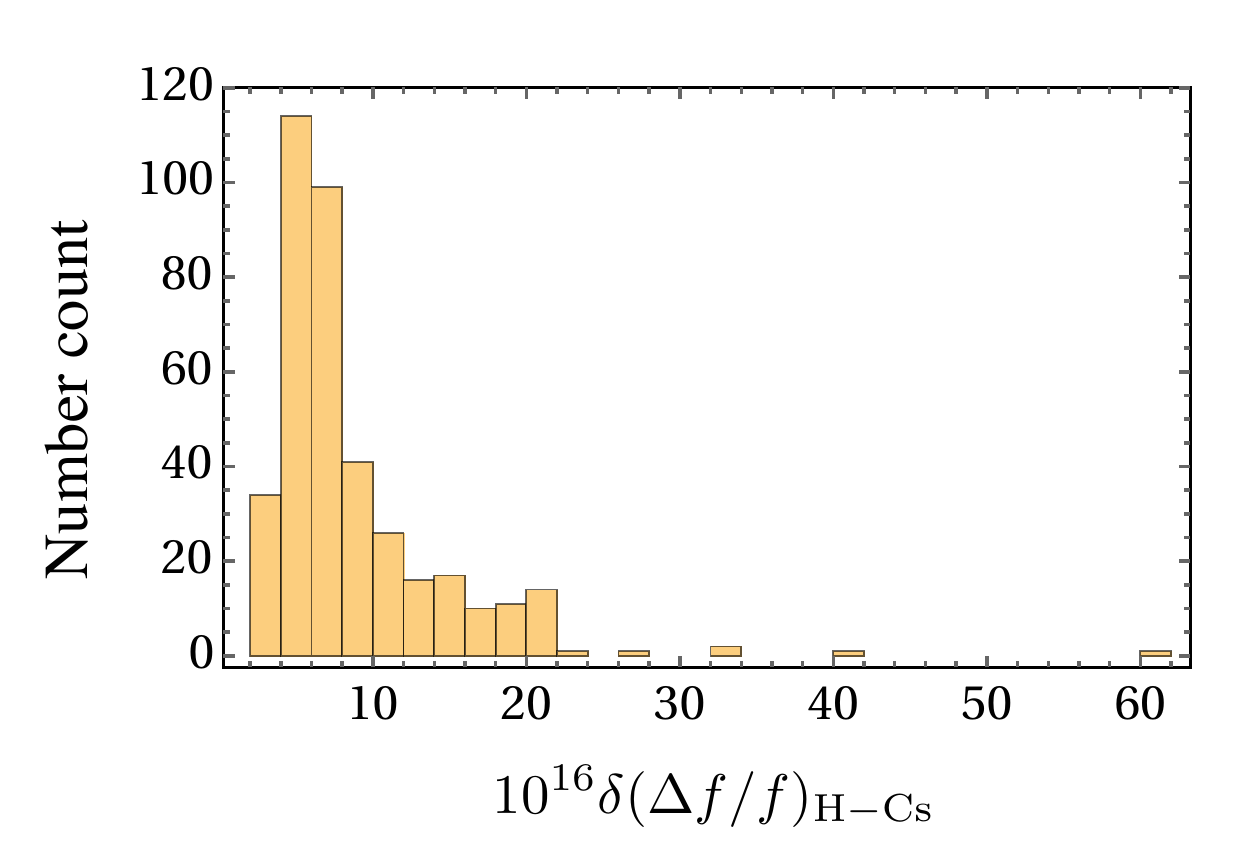}
\caption{Histogram of the fractional frequency difference uncertainties showing a peak population corresponding to $\approx5\times10^{-16}$. The median of the uncertainty before and after MJD54000 is $1.1\times10^{-15}$ compared to $6.35\times10^{-16}$ with 58 and 330 evaluations (number of data points) respectively. All data points before MJD54000 were included in Ashby et al.(2007)\cite{ashby07}.}
\end{center}
\end{figure} 
The final result is obtained by taking the weighted average and adding the uncertainties in quadrature
\begin{equation}
\beta=\sum_{_j}w_j \beta_j=
\left(\sum_{_j}\frac{1}{\delta\beta_j^2}\right)^{-1}\left(\sum_j \frac{\beta_j}{\delta\beta_j^2}\right),
\end{equation}
where $w_j$ are the weights.
The $1\sigma$ uncertainty is obtained by deriving a probability distribution function, for a normal distribution for the residuals, see Fig,~3, from which $\beta_j$ is obtained. We provide the final result for the probability distribution function
\begin{equation}
\mathcal{P}(\beta=\sum w_j \beta_j)=\frac{1}{\sqrt{2\pi}}\left(\frac{1}{\sum_{_j}w_j^2 \delta\beta_j^2}\right)^{1/2}\exp\left(-\frac{\left(\beta-\sum_{_j}w_j \delta\beta_j\right)^2}{2 \sum_{_j}w_j^2 \delta\beta_j^2} \right).
\end{equation}
The $1\sigma$ uncertainty is $\delta\beta=\sqrt{\sum_{_j}w_j^2 \delta\beta_j^2}$.

\begin{acknowledgements}
We acknowledge funding from NASA grant NNH12AT81I. We also thank the atomic standards group at NIST for maintaining the H-masers and sharing the data. We thank Elizabeth Donley, Steven Jefferts, and Chris Oates for providing valuable suggestions that have helped improve this paper.
We thank Yun Ye for discussing the planned clock comparisons between NIST and JILA.
We are very grateful to the anonymous referees for providing valuable suggestion that have helped improve this paper.
\end{acknowledgements}

\bibliography{lpi_rev}

\begin{thebibliography}{42}%
\makeatletter
\providecommand \@ifxundefined [1]{%
 \@ifx{#1\undefined}
}%
\providecommand \@ifnum [1]{%
 \ifnum #1\expandafter \@firstoftwo
 \else \expandafter \@secondoftwo
 \fi
}%
\providecommand \@ifx [1]{%
 \ifx #1\expandafter \@firstoftwo
 \else \expandafter \@secondoftwo
 \fi
}%
\providecommand \natexlab [1]{#1}%
\providecommand \enquote  [1]{``#1''}%
\providecommand \bibnamefont  [1]{#1}%
\providecommand \bibfnamefont [1]{#1}%
\providecommand \citenamefont [1]{#1}%
\providecommand \href@noop [0]{\@secondoftwo}%
\providecommand \href [0]{\begingroup \@sanitize@url \@href}%
\providecommand \@href[1]{\@@startlink{#1}\@@href}%
\providecommand \@@href[1]{\endgroup#1\@@endlink}%
\providecommand \@sanitize@url [0]{\catcode `\\12\catcode `\$12\catcode
  `\&12\catcode `\#12\catcode `\^12\catcode `\_12\catcode `\%12\relax}%
\providecommand \@@startlink[1]{}%
\providecommand \@@endlink[0]{}%
\providecommand \url  [0]{\begingroup\@sanitize@url \@url }%
\providecommand \@url [1]{\endgroup\@href {#1}{\urlprefix }}%
\providecommand \urlprefix  [0]{URL }%
\providecommand \Eprint [0]{\href }%
\providecommand \doibase [0]{http://dx.doi.org/}%
\providecommand \selectlanguage [0]{\@gobble}%
\providecommand \bibinfo  [0]{\@secondoftwo}%
\providecommand \bibfield  [0]{\@secondoftwo}%
\providecommand \translation [1]{[#1]}%
\providecommand \BibitemOpen [0]{}%
\providecommand \bibitemStop [0]{}%
\providecommand \bibitemNoStop [0]{.\EOS\space}%
\providecommand \EOS [0]{\spacefactor3000\relax}%
\providecommand \BibitemShut  [1]{\csname bibitem#1\endcsname}%
\let\auto@bib@innerbib\@empty
\bibitem [{\citenamefont {Einstein}(1996)}]{einstein96}%
  \BibitemOpen
  \bibfield  {author} {\bibinfo {author} {\bibfnamefont {A.}~\bibnamefont
  {Einstein}},\ }\href@noop {} {\emph {\bibinfo {title} {{The Collected Papers
  of Albert Einstein, Volume 6: The Berlin Years: Writings, 1914-1917}}}}\
  (\bibinfo  {publisher} {Princeton University Press},\ \bibinfo {year}
  {1996})\BibitemShut {NoStop}%
\bibitem [{\citenamefont {Weinberg}(1972)}]{weinberg72}%
  \BibitemOpen
  \bibfield  {author} {\bibinfo {author} {\bibfnamefont {S.}~\bibnamefont
  {Weinberg}},\ }\href@noop {} {\emph {\bibinfo {title} {{Gravitation and
  Cosmology}}}}\ (\bibinfo  {publisher} {John Wiley and Sons},\ \bibinfo
  {address} {New York},\ \bibinfo {year} {1972})\BibitemShut {NoStop}%
\bibitem [{\citenamefont {DeMille}\ \emph {et~al.}(2017)\citenamefont
  {DeMille}, \citenamefont {Doyle},\ and\ \citenamefont {Sushkov}}]{demille17}%
  \BibitemOpen
  \bibfield  {author} {\bibinfo {author} {\bibfnamefont {D.}~\bibnamefont
  {DeMille}}, \bibinfo {author} {\bibfnamefont {J.~M.}\ \bibnamefont {Doyle}},
  \ and\ \bibinfo {author} {\bibfnamefont {A.~O.}\ \bibnamefont {Sushkov}},\
  }\href {\doibase 10.1126/science.aal3003} {\bibfield  {journal} {\bibinfo
  {journal} {Science}\ }\textbf {\bibinfo {volume} {357}},\ \bibinfo {pages}
  {990} (\bibinfo {year} {2017})},\ \Eprint {http://arxiv.org/abs/1704.07928}
  {arXiv:1704.07928} \BibitemShut {NoStop}%
\bibitem [{\citenamefont {Will}(2014)}]{Will14}%
  \BibitemOpen
  \bibfield  {author} {\bibinfo {author} {\bibfnamefont {C.~M.}\ \bibnamefont
  {Will}},\ }\href@noop {} {\bibfield  {journal} {\bibinfo  {journal} {Living
  Reviews in Relativity}\ }\textbf {\bibinfo {volume} {17}} (\bibinfo {year}
  {2014})}\BibitemShut {NoStop}%
\bibitem [{\citenamefont {Vessot}\ \emph {et~al.}(1980)\citenamefont {Vessot},
  \citenamefont {Levine}, \citenamefont {Mattison}, \citenamefont {Blomberg},
  \citenamefont {Hoffman}, \citenamefont {Nystrom}, \citenamefont {Farrel},
  \citenamefont {Decher}, \citenamefont {Eby}, \citenamefont {Baugher},
  \citenamefont {Watts}, \citenamefont {Teuber},\ and\ \citenamefont
  {Wills}}]{vessot80}%
  \BibitemOpen
  \bibfield  {author} {\bibinfo {author} {\bibfnamefont {R.~F.}\ \bibnamefont
  {Vessot}}, \bibinfo {author} {\bibfnamefont {M.~W.}\ \bibnamefont {Levine}},
  \bibinfo {author} {\bibfnamefont {E.~M.}\ \bibnamefont {Mattison}}, \bibinfo
  {author} {\bibfnamefont {E.~L.}\ \bibnamefont {Blomberg}}, \bibinfo {author}
  {\bibfnamefont {T.~E.}\ \bibnamefont {Hoffman}}, \bibinfo {author}
  {\bibfnamefont {G.~U.}\ \bibnamefont {Nystrom}}, \bibinfo {author}
  {\bibfnamefont {B.~F.}\ \bibnamefont {Farrel}}, \bibinfo {author}
  {\bibfnamefont {R.}~\bibnamefont {Decher}}, \bibinfo {author} {\bibfnamefont
  {P.~B.}\ \bibnamefont {Eby}}, \bibinfo {author} {\bibfnamefont {C.~R.}\
  \bibnamefont {Baugher}}, \bibinfo {author} {\bibfnamefont {J.~W.}\
  \bibnamefont {Watts}}, \bibinfo {author} {\bibfnamefont {D.~L.}\ \bibnamefont
  {Teuber}}, \ and\ \bibinfo {author} {\bibfnamefont {F.~D.}\ \bibnamefont
  {Wills}},\ }\href {\doibase 10.1103/PhysRevLett.45.2081} {\bibfield
  {journal} {\bibinfo  {journal} {Physical Review Letters}\ }\textbf {\bibinfo
  {volume} {45}},\ \bibinfo {pages} {2081} (\bibinfo {year}
  {1980})}\BibitemShut {NoStop}%
\bibitem [{\citenamefont {{He{\ss}}}\ \emph {et~al.}(2011)\citenamefont
  {{He{\ss}}}, \citenamefont {{Stringhetti}}, \citenamefont {{Hummelsberger}},
  \citenamefont {{Hausner}}, \citenamefont {{Stalford}}, \citenamefont
  {{Nasca}}, \citenamefont {{Cacciapuoti}}, \citenamefont {{Much}},
  \citenamefont {{Feltham}}, \citenamefont {{Vudali}}, \citenamefont
  {{L\'{e}ger}}, \citenamefont {{Picard}}, \citenamefont {{Massonnet}},
  \citenamefont {{Rochat}}, \citenamefont {{Goujon}}, \citenamefont
  {{Sch\"{a}fer}}, \citenamefont {{Laurent}}, \citenamefont {{Lemonde}},
  \citenamefont {{Clairon}}, \citenamefont {{Wolf}}, \citenamefont {{Salomon}},
  \citenamefont {{Proch\'{a}zka}}, \citenamefont {{Schreiber}},\ and\
  \citenamefont {{Montenbruck}}}]{hess11}%
  \BibitemOpen
  \bibfield  {author} {\bibinfo {author} {\bibfnamefont {M.~P.}\ \bibnamefont
  {{He{\ss}}}}, \bibinfo {author} {\bibfnamefont {L.}~\bibnamefont
  {{Stringhetti}}}, \bibinfo {author} {\bibfnamefont {B.}~\bibnamefont
  {{Hummelsberger}}}, \bibinfo {author} {\bibfnamefont {K.}~\bibnamefont
  {{Hausner}}}, \bibinfo {author} {\bibfnamefont {R.}~\bibnamefont
  {{Stalford}}}, \bibinfo {author} {\bibfnamefont {R.}~\bibnamefont {{Nasca}}},
  \bibinfo {author} {\bibfnamefont {L.}~\bibnamefont {{Cacciapuoti}}}, \bibinfo
  {author} {\bibfnamefont {R.}~\bibnamefont {{Much}}}, \bibinfo {author}
  {\bibfnamefont {S.}~\bibnamefont {{Feltham}}}, \bibinfo {author}
  {\bibfnamefont {T.}~\bibnamefont {{Vudali}}}, \bibinfo {author}
  {\bibfnamefont {B.}~\bibnamefont {{L\'{e}ger}}}, \bibinfo {author}
  {\bibfnamefont {F.}~\bibnamefont {{Picard}}}, \bibinfo {author}
  {\bibfnamefont {D.}~\bibnamefont {{Massonnet}}}, \bibinfo {author}
  {\bibfnamefont {P.}~\bibnamefont {{Rochat}}}, \bibinfo {author}
  {\bibfnamefont {D.}~\bibnamefont {{Goujon}}}, \bibinfo {author}
  {\bibfnamefont {W.}~\bibnamefont {{Sch\"{a}fer}}}, \bibinfo {author}
  {\bibfnamefont {P.}~\bibnamefont {{Laurent}}}, \bibinfo {author}
  {\bibfnamefont {P.}~\bibnamefont {{Lemonde}}}, \bibinfo {author}
  {\bibfnamefont {A.}~\bibnamefont {{Clairon}}}, \bibinfo {author}
  {\bibfnamefont {P.}~\bibnamefont {{Wolf}}}, \bibinfo {author} {\bibfnamefont
  {C.}~\bibnamefont {{Salomon}}}, \bibinfo {author} {\bibfnamefont
  {I.}~\bibnamefont {{Proch\'{a}zka}}}, \bibinfo {author} {\bibfnamefont
  {U.}~\bibnamefont {{Schreiber}}}, \ and\ \bibinfo {author} {\bibfnamefont
  {O.}~\bibnamefont {{Montenbruck}}},\ }\href {\doibase
  10.1016/j.actaastro.2011.07.002} {\bibfield  {journal} {\bibinfo  {journal}
  {Acta Astronautica}\ }\textbf {\bibinfo {volume} {69}},\ \bibinfo {pages}
  {929} (\bibinfo {year} {2011})}\BibitemShut {NoStop}%
\bibitem [{\citenamefont {Altschul}\ \emph {et~al.}(2015)\citenamefont
  {Altschul}, \citenamefont {Bailey}, \citenamefont {Blanchet}, \citenamefont
  {Bongs}, \citenamefont {Bouyer}, \citenamefont {Cacciapuoti}, \citenamefont
  {Capozziello}, \citenamefont {Gaaloul}, \citenamefont {Giulini},
  \citenamefont {Hartwig}, \citenamefont {Iess}, \citenamefont {Jetzer},
  \citenamefont {Landragin}, \citenamefont {Rasel}, \citenamefont {Reynaud},
  \citenamefont {Schiller}, \citenamefont {Schubert}, \citenamefont
  {Sorrentino}, \citenamefont {Sterr}, \citenamefont {Tasson}, \citenamefont
  {Tino}, \citenamefont {Tuckey},\ and\ \citenamefont {Wolf}}]{altschul15}%
  \BibitemOpen
  \bibfield  {author} {\bibinfo {author} {\bibfnamefont {B.}~\bibnamefont
  {Altschul}}, \bibinfo {author} {\bibfnamefont {Q.~G.}\ \bibnamefont
  {Bailey}}, \bibinfo {author} {\bibfnamefont {L.}~\bibnamefont {Blanchet}},
  \bibinfo {author} {\bibfnamefont {K.}~\bibnamefont {Bongs}}, \bibinfo
  {author} {\bibfnamefont {P.}~\bibnamefont {Bouyer}}, \bibinfo {author}
  {\bibfnamefont {L.}~\bibnamefont {Cacciapuoti}}, \bibinfo {author}
  {\bibfnamefont {S.}~\bibnamefont {Capozziello}}, \bibinfo {author}
  {\bibfnamefont {N.}~\bibnamefont {Gaaloul}}, \bibinfo {author} {\bibfnamefont
  {D.}~\bibnamefont {Giulini}}, \bibinfo {author} {\bibfnamefont
  {J.}~\bibnamefont {Hartwig}}, \bibinfo {author} {\bibfnamefont
  {L.}~\bibnamefont {Iess}}, \bibinfo {author} {\bibfnamefont {P.}~\bibnamefont
  {Jetzer}}, \bibinfo {author} {\bibfnamefont {A.}~\bibnamefont {Landragin}},
  \bibinfo {author} {\bibfnamefont {E.}~\bibnamefont {Rasel}}, \bibinfo
  {author} {\bibfnamefont {S.}~\bibnamefont {Reynaud}}, \bibinfo {author}
  {\bibfnamefont {S.}~\bibnamefont {Schiller}}, \bibinfo {author}
  {\bibfnamefont {C.}~\bibnamefont {Schubert}}, \bibinfo {author}
  {\bibfnamefont {F.}~\bibnamefont {Sorrentino}}, \bibinfo {author}
  {\bibfnamefont {U.}~\bibnamefont {Sterr}}, \bibinfo {author} {\bibfnamefont
  {J.~D.}\ \bibnamefont {Tasson}}, \bibinfo {author} {\bibfnamefont {G.~M.}\
  \bibnamefont {Tino}}, \bibinfo {author} {\bibfnamefont {P.}~\bibnamefont
  {Tuckey}}, \ and\ \bibinfo {author} {\bibfnamefont {P.}~\bibnamefont
  {Wolf}},\ }\href {\doibase 10.1016/j.asr.2014.07.014} {\bibfield  {journal}
  {\bibinfo  {journal} {Advances in Space Research}\ }\textbf {\bibinfo
  {volume} {55}},\ \bibinfo {pages} {501} (\bibinfo {year} {2015})},\ \Eprint
  {http://arxiv.org/abs/1404.4307} {arXiv:1404.4307} \BibitemShut {NoStop}%
\bibitem [{\citenamefont {Sullivan}(2001)}]{sullivan01}%
  \BibitemOpen
  \bibfield  {author} {\bibinfo {author} {\bibfnamefont {D.}~\bibnamefont
  {Sullivan}},\ }in\ \href {\doibase 10.1109/FREQ.2001.956152} {\emph {\bibinfo
  {booktitle} {{Proceedings of the 2001 IEEE International Frequncy Control
  Symposium and PDA Exhibition (Cat. No.01CH37218)}}}}\ (\bibinfo  {publisher}
  {IEEE},\ \bibinfo {year} {2001})\ pp.\ \bibinfo {pages} {4--17}\BibitemShut
  {NoStop}%
\bibitem [{\citenamefont {Dicke}(1964)}]{dicke64}%
  \BibitemOpen
  \bibfield  {author} {\bibinfo {author} {\bibfnamefont {R.~H.}\ \bibnamefont
  {Dicke}},\ }\href@noop {} {\emph {\bibinfo {title} {{Relativit\'{e}, Groupes
  et Topologie/Relativity, Groups and Topology : Lectures delivered at Les
  Houches during the 1963 session of the summer school of theoretical physics,
  University of Grenoble.{\~{}}Universit\'{e} de Grenoble, Ecole d'\'{e}t\'{e}
  de physique th\'{e}orique, Le}}}}\ (\bibinfo {year} {1964})\BibitemShut
  {NoStop}%
\bibitem [{\citenamefont {Ashby}\ \emph {et~al.}(2007)\citenamefont {Ashby},
  \citenamefont {Heavner}, \citenamefont {Jefferts}, \citenamefont {Parker},
  \citenamefont {Radnaev},\ and\ \citenamefont {Dudin}}]{ashby07}%
  \BibitemOpen
  \bibfield  {author} {\bibinfo {author} {\bibfnamefont {N.}~\bibnamefont
  {Ashby}}, \bibinfo {author} {\bibfnamefont {T.~P.}\ \bibnamefont {Heavner}},
  \bibinfo {author} {\bibfnamefont {S.~R.}\ \bibnamefont {Jefferts}}, \bibinfo
  {author} {\bibfnamefont {T.~E.}\ \bibnamefont {Parker}}, \bibinfo {author}
  {\bibfnamefont {A.~G.}\ \bibnamefont {Radnaev}}, \ and\ \bibinfo {author}
  {\bibfnamefont {Y.~O.}\ \bibnamefont {Dudin}},\ }\href {\doibase
  10.1103/PhysRevLett.98.070802} {\bibfield  {journal} {\bibinfo  {journal}
  {Physical Review Letters}\ }\textbf {\bibinfo {volume} {98}},\ \bibinfo
  {pages} {70802} (\bibinfo {year} {2007})}\BibitemShut {NoStop}%
\bibitem [{\citenamefont {Turneaure}\ \emph {et~al.}(1983)\citenamefont
  {Turneaure}, \citenamefont {Will}, \citenamefont {Farrell}, \citenamefont
  {Mattison},\ and\ \citenamefont {Vessot}}]{turneaure83}%
  \BibitemOpen
  \bibfield  {author} {\bibinfo {author} {\bibfnamefont {J.~P.}\ \bibnamefont
  {Turneaure}}, \bibinfo {author} {\bibfnamefont {C.~M.}\ \bibnamefont {Will}},
  \bibinfo {author} {\bibfnamefont {B.~F.}\ \bibnamefont {Farrell}}, \bibinfo
  {author} {\bibfnamefont {E.~M.}\ \bibnamefont {Mattison}}, \ and\ \bibinfo
  {author} {\bibfnamefont {R.~F.}\ \bibnamefont {Vessot}},\ }\href {\doibase
  10.1103/PhysRevD.27.1705} {\bibfield  {journal} {\bibinfo  {journal}
  {Physical Review D}\ }\textbf {\bibinfo {volume} {27}},\ \bibinfo {pages}
  {1705} (\bibinfo {year} {1983})}\BibitemShut {NoStop}%
\bibitem [{\citenamefont {Godone}\ \emph {et~al.}(1995)\citenamefont {Godone},
  \citenamefont {Novero},\ and\ \citenamefont {Tavella}}]{godone95}%
  \BibitemOpen
  \bibfield  {author} {\bibinfo {author} {\bibfnamefont {A.}~\bibnamefont
  {Godone}}, \bibinfo {author} {\bibfnamefont {C.}~\bibnamefont {Novero}}, \
  and\ \bibinfo {author} {\bibfnamefont {P.}~\bibnamefont {Tavella}},\ }\href
  {\doibase 10.1103/PhysRevD.51.319} {\bibfield  {journal} {\bibinfo  {journal}
  {Physical Review D}\ }\textbf {\bibinfo {volume} {51}},\ \bibinfo {pages}
  {319} (\bibinfo {year} {1995})}\BibitemShut {NoStop}%
\bibitem [{\citenamefont {Gu\'{e}na}\ \emph {et~al.}(2012)\citenamefont
  {Gu\'{e}na}, \citenamefont {Abgrall}, \citenamefont {Rovera}, \citenamefont
  {Rosenbusch}, \citenamefont {Tobar}, \citenamefont {Laurent}, \citenamefont
  {Clairon},\ and\ \citenamefont {Bize}}]{guena_dada_12}%
  \BibitemOpen
  \bibfield  {author} {\bibinfo {author} {\bibfnamefont {J.}~\bibnamefont
  {Gu\'{e}na}}, \bibinfo {author} {\bibfnamefont {M.}~\bibnamefont {Abgrall}},
  \bibinfo {author} {\bibfnamefont {D.}~\bibnamefont {Rovera}}, \bibinfo
  {author} {\bibfnamefont {P.}~\bibnamefont {Rosenbusch}}, \bibinfo {author}
  {\bibfnamefont {M.~E.}\ \bibnamefont {Tobar}}, \bibinfo {author}
  {\bibfnamefont {P.}~\bibnamefont {Laurent}}, \bibinfo {author} {\bibfnamefont
  {A.}~\bibnamefont {Clairon}}, \ and\ \bibinfo {author} {\bibfnamefont
  {S.}~\bibnamefont {Bize}},\ }\href {\doibase 10.1103/PhysRevLett.109.080801}
  {\bibfield  {journal} {\bibinfo  {journal} {Physical Review Letters}\
  }\textbf {\bibinfo {volume} {109}},\ \bibinfo {pages} {80801} (\bibinfo
  {year} {2012})},\ \Eprint {http://arxiv.org/abs/1205.4235} {arXiv:1205.4235}
  \BibitemShut {NoStop}%
\bibitem [{\citenamefont {Peil}\ \emph {et~al.}(2013)\citenamefont {Peil},
  \citenamefont {Crane}, \citenamefont {Hanssen}, \citenamefont {Swanson},\
  and\ \citenamefont {Ekstrom}}]{peil13}%
  \BibitemOpen
  \bibfield  {author} {\bibinfo {author} {\bibfnamefont {S.}~\bibnamefont
  {Peil}}, \bibinfo {author} {\bibfnamefont {S.}~\bibnamefont {Crane}},
  \bibinfo {author} {\bibfnamefont {J.~L.}\ \bibnamefont {Hanssen}}, \bibinfo
  {author} {\bibfnamefont {T.~B.}\ \bibnamefont {Swanson}}, \ and\ \bibinfo
  {author} {\bibfnamefont {C.~R.}\ \bibnamefont {Ekstrom}},\ }\href {\doibase
  10.1103/PhysRevA.87.010102} {\bibfield  {journal} {\bibinfo  {journal}
  {Physical Review A - Atomic, Molecular, and Optical Physics}\ }\textbf
  {\bibinfo {volume} {87}} (\bibinfo {year} {2013}),\
  10.1103/PhysRevA.87.010102},\ \Eprint {http://arxiv.org/abs/1301.6145v1}
  {arXiv:1301.6145v1 [physics.atom-ph]} \BibitemShut {NoStop}%
\bibitem [{\citenamefont {Levi}\ \emph {et~al.}(2014)\citenamefont {Levi},
  \citenamefont {Calonico}, \citenamefont {Calosso}, \citenamefont {Godone},
  \citenamefont {Micalizio},\ and\ \citenamefont {Costanzo}}]{torinof1}%
  \BibitemOpen
  \bibfield  {author} {\bibinfo {author} {\bibfnamefont {F.}~\bibnamefont
  {Levi}}, \bibinfo {author} {\bibfnamefont {D.}~\bibnamefont {Calonico}},
  \bibinfo {author} {\bibfnamefont {C.~E.}\ \bibnamefont {Calosso}}, \bibinfo
  {author} {\bibfnamefont {A.}~\bibnamefont {Godone}}, \bibinfo {author}
  {\bibfnamefont {S.}~\bibnamefont {Micalizio}}, \ and\ \bibinfo {author}
  {\bibfnamefont {G.~A.}\ \bibnamefont {Costanzo}},\ }\href {\doibase
  10.1088/0026-1394/51/3/270} {\bibfield  {journal} {\bibinfo  {journal}
  {Metrologia}\ }\textbf {\bibinfo {volume} {51}},\ \bibinfo {pages} {270}
  (\bibinfo {year} {2014})}\BibitemShut {NoStop}%
\bibitem [{\citenamefont {Jefferts}\ \emph {et~al.}(2003)\citenamefont
  {Jefferts}, \citenamefont {Shirley}, \citenamefont {Parker}, \citenamefont
  {Heavner}, \citenamefont {Meekhof}, \citenamefont {Nelson}, \citenamefont
  {Levi}, \citenamefont {Costanzo}, \citenamefont {Marchi}, \citenamefont
  {Drullinger}, \citenamefont {Hollberg}, \citenamefont {Lee},\ and\
  \citenamefont {Walls}}]{nistf1}%
  \BibitemOpen
  \bibfield  {author} {\bibinfo {author} {\bibfnamefont {S.~R.}\ \bibnamefont
  {Jefferts}}, \bibinfo {author} {\bibfnamefont {J.}~\bibnamefont {Shirley}},
  \bibinfo {author} {\bibfnamefont {T.~E.}\ \bibnamefont {Parker}}, \bibinfo
  {author} {\bibfnamefont {T.~P.}\ \bibnamefont {Heavner}}, \bibinfo {author}
  {\bibfnamefont {D.~M.}\ \bibnamefont {Meekhof}}, \bibinfo {author}
  {\bibfnamefont {C.}~\bibnamefont {Nelson}}, \bibinfo {author} {\bibfnamefont
  {F.}~\bibnamefont {Levi}}, \bibinfo {author} {\bibfnamefont {G.}~\bibnamefont
  {Costanzo}}, \bibinfo {author} {\bibfnamefont {a.~D.}\ \bibnamefont
  {Marchi}}, \bibinfo {author} {\bibfnamefont {R.}~\bibnamefont {Drullinger}},
  \bibinfo {author} {\bibfnamefont {L.}~\bibnamefont {Hollberg}}, \bibinfo
  {author} {\bibfnamefont {W.~D.}\ \bibnamefont {Lee}}, \ and\ \bibinfo
  {author} {\bibfnamefont {F.~L.}\ \bibnamefont {Walls}},\ }\href {\doibase
  10.1088/0026-1394/39/4/1} {\bibfield  {journal} {\bibinfo  {journal}
  {Metrologia}\ }\textbf {\bibinfo {volume} {39}},\ \bibinfo {pages} {321}
  (\bibinfo {year} {2003})}\BibitemShut {NoStop}%
\bibitem [{\citenamefont {Weyers}\ \emph {et~al.}(2001)\citenamefont {Weyers},
  \citenamefont {Hubner}, \citenamefont {Schroder}, \citenamefont {Tamm},\ and\
  \citenamefont {Bauch}}]{ptbf1}%
  \BibitemOpen
  \bibfield  {author} {\bibinfo {author} {\bibfnamefont {S.}~\bibnamefont
  {Weyers}}, \bibinfo {author} {\bibfnamefont {U.}~\bibnamefont {Hubner}},
  \bibinfo {author} {\bibfnamefont {R.}~\bibnamefont {Schroder}}, \bibinfo
  {author} {\bibfnamefont {C.}~\bibnamefont {Tamm}}, \ and\ \bibinfo {author}
  {\bibfnamefont {A.}~\bibnamefont {Bauch}},\ }\href {\doibase
  10.1088/0026-1394/38/4/7} {\bibfield  {journal} {\bibinfo  {journal}
  {Metrologia}\ }\textbf {\bibinfo {volume} {38}},\ \bibinfo {pages} {343}
  (\bibinfo {year} {2001})}\BibitemShut {NoStop}%
\bibitem [{\citenamefont {Gerginov}\ \emph {et~al.}(2010)\citenamefont
  {Gerginov}, \citenamefont {Nemitz}, \citenamefont {Weyers}, \citenamefont
  {Schr\"{o}der}, \citenamefont {Griebsch},\ and\ \citenamefont
  {Wynands}}]{ptbf2}%
  \BibitemOpen
  \bibfield  {author} {\bibinfo {author} {\bibfnamefont {V.}~\bibnamefont
  {Gerginov}}, \bibinfo {author} {\bibfnamefont {N.}~\bibnamefont {Nemitz}},
  \bibinfo {author} {\bibfnamefont {S.}~\bibnamefont {Weyers}}, \bibinfo
  {author} {\bibfnamefont {R.}~\bibnamefont {Schr\"{o}der}}, \bibinfo {author}
  {\bibfnamefont {D.}~\bibnamefont {Griebsch}}, \ and\ \bibinfo {author}
  {\bibfnamefont {R.}~\bibnamefont {Wynands}},\ }\href {\doibase
  10.1088/0026-1394/47/1/008} {\bibfield  {journal} {\bibinfo  {journal}
  {Metrologia}\ }\textbf {\bibinfo {volume} {47}},\ \bibinfo {pages} {65}
  (\bibinfo {year} {2010})}\BibitemShut {NoStop}%
\bibitem [{\citenamefont {Szymaniec}\ \emph {et~al.}(2016)\citenamefont
  {Szymaniec}, \citenamefont {Lea}, \citenamefont {Gibble}, \citenamefont
  {Park}, \citenamefont {Liu},\ and\ \citenamefont {G{\l}owacki}}]{nplf}%
  \BibitemOpen
  \bibfield  {author} {\bibinfo {author} {\bibfnamefont {K.}~\bibnamefont
  {Szymaniec}}, \bibinfo {author} {\bibfnamefont {S.~N.}\ \bibnamefont {Lea}},
  \bibinfo {author} {\bibfnamefont {K.}~\bibnamefont {Gibble}}, \bibinfo
  {author} {\bibfnamefont {S.~E.}\ \bibnamefont {Park}}, \bibinfo {author}
  {\bibfnamefont {K.}~\bibnamefont {Liu}}, \ and\ \bibinfo {author}
  {\bibfnamefont {P.}~\bibnamefont {G{\l}owacki}},\ }\href {\doibase
  10.1088/1742-6596/723/1/012003} {\bibfield  {journal} {\bibinfo  {journal}
  {Journal of Physics: Conference Series}\ }\textbf {\bibinfo {volume} {723}},\
  \bibinfo {pages} {012003} (\bibinfo {year} {2016})}\BibitemShut {NoStop}%
\bibitem [{\citenamefont {Guena}\ \emph {et~al.}(2012)\citenamefont {Guena},
  \citenamefont {Abgrall}, \citenamefont {Rovera}, \citenamefont {Laurent},
  \citenamefont {Chupin}, \citenamefont {Lours}, \citenamefont {Santarelli},
  \citenamefont {Rosenbusch}, \citenamefont {Tobar}, \citenamefont {Li},
  \citenamefont {Gibble}, \citenamefont {Clairon},\ and\ \citenamefont
  {Bize}}]{syrtef}%
  \BibitemOpen
  \bibfield  {author} {\bibinfo {author} {\bibfnamefont {J.}~\bibnamefont
  {Guena}}, \bibinfo {author} {\bibfnamefont {M.}~\bibnamefont {Abgrall}},
  \bibinfo {author} {\bibfnamefont {D.}~\bibnamefont {Rovera}}, \bibinfo
  {author} {\bibfnamefont {P.}~\bibnamefont {Laurent}}, \bibinfo {author}
  {\bibfnamefont {B.}~\bibnamefont {Chupin}}, \bibinfo {author} {\bibfnamefont
  {M.}~\bibnamefont {Lours}}, \bibinfo {author} {\bibfnamefont
  {G.}~\bibnamefont {Santarelli}}, \bibinfo {author} {\bibfnamefont
  {P.}~\bibnamefont {Rosenbusch}}, \bibinfo {author} {\bibfnamefont {M.~E.}\
  \bibnamefont {Tobar}}, \bibinfo {author} {\bibfnamefont {R.}~\bibnamefont
  {Li}}, \bibinfo {author} {\bibfnamefont {K.}~\bibnamefont {Gibble}}, \bibinfo
  {author} {\bibfnamefont {A.}~\bibnamefont {Clairon}}, \ and\ \bibinfo
  {author} {\bibfnamefont {S.}~\bibnamefont {Bize}},\ }in\ \href {\doibase
  10.1109/TUFFC.2012.2208} {\emph {\bibinfo {booktitle} {{IEEE Transactions on
  Ultrasonics, Ferroelectrics, and Frequency Control}}}},\ Vol.~\bibinfo
  {volume} {59}\ (\bibinfo {year} {2012})\ pp.\ \bibinfo {pages} {391--410},\
  \Eprint {http://arxiv.org/abs/1204.3621} {arXiv:1204.3621} \BibitemShut
  {NoStop}%
\bibitem [{\citenamefont {BIPM}()}]{bipm}%
  \BibitemOpen
  \bibfield  {author} {\bibinfo {author} {\bibnamefont {BIPM}},\ }\href
  {http://www.bipm.org/en/bipm-services/timescales/time-ftp/data.html}
  {\enquote {\bibinfo {title} {{Reports of evaluation of Primary Frequency
  Standards}},}\ }\bibinfo {howpublished}
  {http://www.bipm.org/en/bipm-services/timescales/time-ftp/data.html}\BibitemShut
  {NoStop}%
\bibitem [{\citenamefont {Parker}(1999)}]{parker99}%
  \BibitemOpen
  \bibfield  {author} {\bibinfo {author} {\bibfnamefont {T.~E.}\ \bibnamefont
  {Parker}},\ }\href {\doibase 10.1109/58.764861} {\bibfield  {journal}
  {\bibinfo  {journal} {IEEE Transactions on Ultrasonics, Ferroelectrics, and
  Frequency Control}\ }\textbf {\bibinfo {volume} {46}},\ \bibinfo {pages}
  {745} (\bibinfo {year} {1999})}\BibitemShut {NoStop}%
\bibitem [{\citenamefont {Parker}\ \emph {et~al.}(2005)\citenamefont {Parker},
  \citenamefont {Jefferts}, \citenamefont {Heavner},\ and\ \citenamefont
  {Donley}}]{parker05}%
  \BibitemOpen
  \bibfield  {author} {\bibinfo {author} {\bibfnamefont {T.~E.}\ \bibnamefont
  {Parker}}, \bibinfo {author} {\bibfnamefont {S.~R.}\ \bibnamefont
  {Jefferts}}, \bibinfo {author} {\bibfnamefont {T.~P.}\ \bibnamefont
  {Heavner}}, \ and\ \bibinfo {author} {\bibfnamefont {E.~A.}\ \bibnamefont
  {Donley}},\ }\href {\doibase 10.1088/0026-1394/42/5/013} {\bibfield
  {journal} {\bibinfo  {journal} {Metrologia}\ }\textbf {\bibinfo {volume}
  {42}},\ \bibinfo {pages} {423} (\bibinfo {year} {2005})}\BibitemShut
  {NoStop}%
\bibitem [{\citenamefont {Lewis}(1991)}]{lewis91}%
  \BibitemOpen
  \bibfield  {author} {\bibinfo {author} {\bibfnamefont {L.~L.}\ \bibnamefont
  {Lewis}},\ }\href {\doibase 10.1109/5.84969} {\bibfield  {journal} {\bibinfo
  {journal} {Proceedings of the IEEE}\ }\textbf {\bibinfo {volume} {79}},\
  \bibinfo {pages} {927} (\bibinfo {year} {1991})}\BibitemShut {NoStop}%
\bibitem [{\citenamefont {Folkner}\ \emph {et~al.}(2014)\citenamefont
  {Folkner}, \citenamefont {Williams}, \citenamefont {Boggs}, \citenamefont
  {Park},\ and\ \citenamefont {Kuchynka}}]{folkner14}%
  \BibitemOpen
  \bibfield  {author} {\bibinfo {author} {\bibfnamefont {W.~M.}\ \bibnamefont
  {Folkner}}, \bibinfo {author} {\bibfnamefont {J.~G.}\ \bibnamefont
  {Williams}}, \bibinfo {author} {\bibfnamefont {D.~H.}\ \bibnamefont {Boggs}},
  \bibinfo {author} {\bibfnamefont {R.~S.}\ \bibnamefont {Park}}, \ and\
  \bibinfo {author} {\bibfnamefont {P.}~\bibnamefont {Kuchynka}},\ }\href@noop
  {} {\bibfield  {journal} {\bibinfo  {journal} {Interplanet. Netw. Prog. Rep}\
  }\textbf {\bibinfo {volume} {196}},\ \bibinfo {pages} {42} (\bibinfo {year}
  {2014})}\BibitemShut {NoStop}%
\bibitem [{\citenamefont {Uzan}(2003)}]{uzan03}%
  \BibitemOpen
  \bibfield  {author} {\bibinfo {author} {\bibfnamefont {J.-P.}\ \bibnamefont
  {Uzan}},\ }\href {\doibase 10.1103/RevModPhys.75.403} {\bibfield  {journal}
  {\bibinfo  {journal} {Rev. Mod. Phys.}\ }\textbf {\bibinfo {volume} {75}},\
  \bibinfo {pages} {403} (\bibinfo {year} {2003})}\BibitemShut {NoStop}%
\bibitem [{\citenamefont {Fischer}\ \emph {et~al.}(2004)\citenamefont
  {Fischer}, \citenamefont {Kolachevsky}, \citenamefont {Zimmermann},
  \citenamefont {Holzwarth}, \citenamefont {Udem}, \citenamefont {H\"{a}nsch},
  \citenamefont {Abgrall}, \citenamefont {Gr\"{u}nert}, \citenamefont
  {Maksimovic}, \citenamefont {Bize}, \citenamefont {Marion}, \citenamefont
  {{Dos Santos}}, \citenamefont {Lemonde}, \citenamefont {Santarelli},
  \citenamefont {Laurent}, \citenamefont {Clairon}, \citenamefont {Salomon},
  \citenamefont {Haas}, \citenamefont {Jentschura},\ and\ \citenamefont
  {Keitel}}]{fischer_alpha_04}%
  \BibitemOpen
  \bibfield  {author} {\bibinfo {author} {\bibfnamefont {M.}~\bibnamefont
  {Fischer}}, \bibinfo {author} {\bibfnamefont {N.}~\bibnamefont
  {Kolachevsky}}, \bibinfo {author} {\bibfnamefont {M.}~\bibnamefont
  {Zimmermann}}, \bibinfo {author} {\bibfnamefont {R.}~\bibnamefont
  {Holzwarth}}, \bibinfo {author} {\bibfnamefont {T.}~\bibnamefont {Udem}},
  \bibinfo {author} {\bibfnamefont {T.~W.}\ \bibnamefont {H\"{a}nsch}},
  \bibinfo {author} {\bibfnamefont {M.}~\bibnamefont {Abgrall}}, \bibinfo
  {author} {\bibfnamefont {J.}~\bibnamefont {Gr\"{u}nert}}, \bibinfo {author}
  {\bibfnamefont {I.}~\bibnamefont {Maksimovic}}, \bibinfo {author}
  {\bibfnamefont {S.}~\bibnamefont {Bize}}, \bibinfo {author} {\bibfnamefont
  {H.}~\bibnamefont {Marion}}, \bibinfo {author} {\bibfnamefont {F.~P.}\
  \bibnamefont {{Dos Santos}}}, \bibinfo {author} {\bibfnamefont
  {P.}~\bibnamefont {Lemonde}}, \bibinfo {author} {\bibfnamefont
  {G.}~\bibnamefont {Santarelli}}, \bibinfo {author} {\bibfnamefont
  {P.}~\bibnamefont {Laurent}}, \bibinfo {author} {\bibfnamefont
  {A.}~\bibnamefont {Clairon}}, \bibinfo {author} {\bibfnamefont
  {C.}~\bibnamefont {Salomon}}, \bibinfo {author} {\bibfnamefont
  {M.}~\bibnamefont {Haas}}, \bibinfo {author} {\bibfnamefont {U.~D.}\
  \bibnamefont {Jentschura}}, \ and\ \bibinfo {author} {\bibfnamefont {C.~H.}\
  \bibnamefont {Keitel}},\ }\href {\doibase 10.1103/PhysRevLett.92.230802}
  {\bibfield  {journal} {\bibinfo  {journal} {Physical Review Letters}\
  }\textbf {\bibinfo {volume} {92}},\ \bibinfo {pages} {230802} (\bibinfo
  {year} {2004})},\ \Eprint {http://arxiv.org/abs/0312086} {arXiv:0312086
  [physics]} \BibitemShut {NoStop}%
\bibitem [{\citenamefont {Flambaum}\ and\ \citenamefont
  {Tedesco}(2006)}]{flambaum06}%
  \BibitemOpen
  \bibfield  {author} {\bibinfo {author} {\bibfnamefont {V.~V.}\ \bibnamefont
  {Flambaum}}\ and\ \bibinfo {author} {\bibfnamefont {A.~F.}\ \bibnamefont
  {Tedesco}},\ }\href {\doibase 10.1103/PhysRevC.73.055501} {\bibfield
  {journal} {\bibinfo  {journal} {Phys. Rev. C}\ }\textbf {\bibinfo {volume}
  {73}},\ \bibinfo {pages} {55501} (\bibinfo {year} {2006})}\BibitemShut
  {NoStop}%
\bibitem [{\citenamefont {Dinh}\ \emph {et~al.}(2009)\citenamefont {Dinh},
  \citenamefont {Dunning}, \citenamefont {Dzuba},\ and\ \citenamefont
  {Flambaum}}]{dinh_quark_09}%
  \BibitemOpen
  \bibfield  {author} {\bibinfo {author} {\bibfnamefont {T.~H.}\ \bibnamefont
  {Dinh}}, \bibinfo {author} {\bibfnamefont {A.}~\bibnamefont {Dunning}},
  \bibinfo {author} {\bibfnamefont {V.~A.}\ \bibnamefont {Dzuba}}, \ and\
  \bibinfo {author} {\bibfnamefont {V.~V.}\ \bibnamefont {Flambaum}},\ }\href
  {\doibase 10.1103/PhysRevA.79.054102} {\bibfield  {journal} {\bibinfo
  {journal} {Physical Review A - Atomic, Molecular, and Optical Physics}\
  }\textbf {\bibinfo {volume} {79}},\ \bibinfo {pages} {54102} (\bibinfo {year}
  {2009})},\ \Eprint {http://arxiv.org/abs/0903.2090} {arXiv:0903.2090}
  \BibitemShut {NoStop}%
\bibitem [{\citenamefont {Godun}\ \emph {et~al.}(2014)\citenamefont {Godun},
  \citenamefont {Nisbet-Jones}, \citenamefont {Jones}, \citenamefont {King},
  \citenamefont {Johnson}, \citenamefont {Margolis}, \citenamefont {Szymaniec},
  \citenamefont {Lea}, \citenamefont {Bongs},\ and\ \citenamefont
  {Gill}}]{godun14}%
  \BibitemOpen
  \bibfield  {author} {\bibinfo {author} {\bibfnamefont {R.~M.}\ \bibnamefont
  {Godun}}, \bibinfo {author} {\bibfnamefont {P.~B.~R.}\ \bibnamefont
  {Nisbet-Jones}}, \bibinfo {author} {\bibfnamefont {J.~M.}\ \bibnamefont
  {Jones}}, \bibinfo {author} {\bibfnamefont {S.~A.}\ \bibnamefont {King}},
  \bibinfo {author} {\bibfnamefont {L.~A.~M.}\ \bibnamefont {Johnson}},
  \bibinfo {author} {\bibfnamefont {H.~S.}\ \bibnamefont {Margolis}}, \bibinfo
  {author} {\bibfnamefont {K.}~\bibnamefont {Szymaniec}}, \bibinfo {author}
  {\bibfnamefont {S.~N.}\ \bibnamefont {Lea}}, \bibinfo {author} {\bibfnamefont
  {K.}~\bibnamefont {Bongs}}, \ and\ \bibinfo {author} {\bibfnamefont
  {P.}~\bibnamefont {Gill}},\ }\href {\doibase 10.1103/PhysRevLett.113.210801}
  {\bibfield  {journal} {\bibinfo  {journal} {Phys. Rev. Lett.}\ }\textbf
  {\bibinfo {volume} {113}},\ \bibinfo {pages} {210801} (\bibinfo {year}
  {2014})}\BibitemShut {NoStop}%
\bibitem [{\citenamefont {Huntemann}\ \emph {et~al.}(2014)\citenamefont
  {Huntemann}, \citenamefont {Lipphardt}, \citenamefont {Tamm}, \citenamefont
  {Gerginov}, \citenamefont {Weyers},\ and\ \citenamefont {Peik}}]{hunt14}%
  \BibitemOpen
  \bibfield  {author} {\bibinfo {author} {\bibfnamefont {N.}~\bibnamefont
  {Huntemann}}, \bibinfo {author} {\bibfnamefont {B.}~\bibnamefont
  {Lipphardt}}, \bibinfo {author} {\bibfnamefont {C.}~\bibnamefont {Tamm}},
  \bibinfo {author} {\bibfnamefont {V.}~\bibnamefont {Gerginov}}, \bibinfo
  {author} {\bibfnamefont {S.}~\bibnamefont {Weyers}}, \ and\ \bibinfo {author}
  {\bibfnamefont {E.}~\bibnamefont {Peik}},\ }\href {\doibase
  10.1103/PhysRevLett.113.210802} {\bibfield  {journal} {\bibinfo  {journal}
  {Phys. Rev. Lett.}\ }\textbf {\bibinfo {volume} {113}},\ \bibinfo {pages}
  {210802} (\bibinfo {year} {2014})}\BibitemShut {NoStop}%
\bibitem [{\citenamefont {Ludlow}\ \emph {et~al.}(2015)\citenamefont {Ludlow},
  \citenamefont {Boyd}, \citenamefont {Ye}, \citenamefont {Peik},\ and\
  \citenamefont {Schmidt}}]{ludlow15}%
  \BibitemOpen
  \bibfield  {author} {\bibinfo {author} {\bibfnamefont {A.~D.}\ \bibnamefont
  {Ludlow}}, \bibinfo {author} {\bibfnamefont {M.~M.}\ \bibnamefont {Boyd}},
  \bibinfo {author} {\bibfnamefont {J.}~\bibnamefont {Ye}}, \bibinfo {author}
  {\bibfnamefont {E.}~\bibnamefont {Peik}}, \ and\ \bibinfo {author}
  {\bibfnamefont {P.~O.}\ \bibnamefont {Schmidt}},\ }\href {\doibase
  10.1103/RevModPhys.87.637} {\bibfield  {journal} {\bibinfo  {journal} {Rev.
  Mod. Phys.}\ }\textbf {\bibinfo {volume} {87}},\ \bibinfo {pages} {637}
  (\bibinfo {year} {2015})}\BibitemShut {NoStop}%
\bibitem [{\citenamefont {Lisdat}\ \emph {et~al.}(2016)\citenamefont {Lisdat},
  \citenamefont {Grosche}, \citenamefont {Quintin}, \citenamefont {Shi},
  \citenamefont {Raupach}, \citenamefont {Grebing}, \citenamefont {Nicolodi},
  \citenamefont {Stefani}, \citenamefont {Al-Masoudi}, \citenamefont
  {D\"{o}rscher}, \citenamefont {H\"{a}fner}, \citenamefont {Robyr},
  \citenamefont {Chiodo}, \citenamefont {Bilicki}, \citenamefont {Bookjans},
  \citenamefont {Koczwara}, \citenamefont {Koke}, \citenamefont {Kuhl},
  \citenamefont {Wiotte}, \citenamefont {Meynadier}, \citenamefont {Camisard},
  \citenamefont {Abgrall}, \citenamefont {Lours}, \citenamefont {Legero},
  \citenamefont {Schnatz}, \citenamefont {Sterr}, \citenamefont {Denker},
  \citenamefont {Chardonnet}, \citenamefont {{Le Coq}}, \citenamefont
  {Santarelli}, \citenamefont {Amy-Klein}, \citenamefont {{Le Targat}},
  \citenamefont {Lodewyck}, \citenamefont {Lopez},\ and\ \citenamefont
  {Pottie}}]{lisdat16}%
  \BibitemOpen
  \bibfield  {author} {\bibinfo {author} {\bibfnamefont {C.}~\bibnamefont
  {Lisdat}}, \bibinfo {author} {\bibfnamefont {G.}~\bibnamefont {Grosche}},
  \bibinfo {author} {\bibfnamefont {N.}~\bibnamefont {Quintin}}, \bibinfo
  {author} {\bibfnamefont {C.}~\bibnamefont {Shi}}, \bibinfo {author}
  {\bibfnamefont {S.}~\bibnamefont {Raupach}}, \bibinfo {author} {\bibfnamefont
  {C.}~\bibnamefont {Grebing}}, \bibinfo {author} {\bibfnamefont
  {D.}~\bibnamefont {Nicolodi}}, \bibinfo {author} {\bibfnamefont
  {F.}~\bibnamefont {Stefani}}, \bibinfo {author} {\bibfnamefont
  {A.}~\bibnamefont {Al-Masoudi}}, \bibinfo {author} {\bibfnamefont
  {S.}~\bibnamefont {D\"{o}rscher}}, \bibinfo {author} {\bibfnamefont
  {S.}~\bibnamefont {H\"{a}fner}}, \bibinfo {author} {\bibfnamefont {J.-L.}\
  \bibnamefont {Robyr}}, \bibinfo {author} {\bibfnamefont {N.}~\bibnamefont
  {Chiodo}}, \bibinfo {author} {\bibfnamefont {S.}~\bibnamefont {Bilicki}},
  \bibinfo {author} {\bibfnamefont {E.}~\bibnamefont {Bookjans}}, \bibinfo
  {author} {\bibfnamefont {A.}~\bibnamefont {Koczwara}}, \bibinfo {author}
  {\bibfnamefont {S.}~\bibnamefont {Koke}}, \bibinfo {author} {\bibfnamefont
  {A.}~\bibnamefont {Kuhl}}, \bibinfo {author} {\bibfnamefont {F.}~\bibnamefont
  {Wiotte}}, \bibinfo {author} {\bibfnamefont {F.}~\bibnamefont {Meynadier}},
  \bibinfo {author} {\bibfnamefont {E.}~\bibnamefont {Camisard}}, \bibinfo
  {author} {\bibfnamefont {M.}~\bibnamefont {Abgrall}}, \bibinfo {author}
  {\bibfnamefont {M.}~\bibnamefont {Lours}}, \bibinfo {author} {\bibfnamefont
  {T.}~\bibnamefont {Legero}}, \bibinfo {author} {\bibfnamefont
  {H.}~\bibnamefont {Schnatz}}, \bibinfo {author} {\bibfnamefont
  {U.}~\bibnamefont {Sterr}}, \bibinfo {author} {\bibfnamefont
  {H.}~\bibnamefont {Denker}}, \bibinfo {author} {\bibfnamefont
  {C.}~\bibnamefont {Chardonnet}}, \bibinfo {author} {\bibfnamefont
  {Y.}~\bibnamefont {{Le Coq}}}, \bibinfo {author} {\bibfnamefont
  {G.}~\bibnamefont {Santarelli}}, \bibinfo {author} {\bibfnamefont
  {A.}~\bibnamefont {Amy-Klein}}, \bibinfo {author} {\bibfnamefont
  {R.}~\bibnamefont {{Le Targat}}}, \bibinfo {author} {\bibfnamefont
  {J.}~\bibnamefont {Lodewyck}}, \bibinfo {author} {\bibfnamefont
  {O.}~\bibnamefont {Lopez}}, \ and\ \bibinfo {author} {\bibfnamefont {P.-E.}\
  \bibnamefont {Pottie}},\ }\href {\doibase 10.1038/ncomms12443} {\bibfield
  {journal} {\bibinfo  {journal} {Nature Communications}\ }\textbf {\bibinfo
  {volume} {7}},\ \bibinfo {pages} {12443} (\bibinfo {year} {2016})},\ \Eprint
  {http://arxiv.org/abs/1511.07735} {arXiv:1511.07735} \BibitemShut {NoStop}%
\bibitem [{\citenamefont {Bloom}\ \emph {et~al.}(2014)\citenamefont {Bloom},
  \citenamefont {Nicholson}, \citenamefont {Williams}, \citenamefont
  {Campbell}, \citenamefont {Bishof}, \citenamefont {Zhang}, \citenamefont
  {Zhang}, \citenamefont {Bromley},\ and\ \citenamefont {Ye}}]{bloom14}%
  \BibitemOpen
  \bibfield  {author} {\bibinfo {author} {\bibfnamefont {B.~J.}\ \bibnamefont
  {Bloom}}, \bibinfo {author} {\bibfnamefont {T.~L.}\ \bibnamefont
  {Nicholson}}, \bibinfo {author} {\bibfnamefont {J.~R.}\ \bibnamefont
  {Williams}}, \bibinfo {author} {\bibfnamefont {S.~L.}\ \bibnamefont
  {Campbell}}, \bibinfo {author} {\bibfnamefont {M.}~\bibnamefont {Bishof}},
  \bibinfo {author} {\bibfnamefont {X.}~\bibnamefont {Zhang}}, \bibinfo
  {author} {\bibfnamefont {W.}~\bibnamefont {Zhang}}, \bibinfo {author}
  {\bibfnamefont {S.~L.}\ \bibnamefont {Bromley}}, \ and\ \bibinfo {author}
  {\bibfnamefont {J.}~\bibnamefont {Ye}},\ }\href {\doibase
  10.1038/nature12941} {\bibfield  {journal} {\bibinfo  {journal} {Nature}\
  }\textbf {\bibinfo {volume} {506}},\ \bibinfo {pages} {71} (\bibinfo {year}
  {2014})},\ \Eprint {http://arxiv.org/abs/1309.1137} {arXiv:1309.1137}
  \BibitemShut {NoStop}%
\bibitem [{\citenamefont {Nicholson}\ \emph {et~al.}(2015)\citenamefont
  {Nicholson}, \citenamefont {Campbell}, \citenamefont {Hutson}, \citenamefont
  {Marti}, \citenamefont {Bloom}, \citenamefont {McNally}, \citenamefont
  {Zhang}, \citenamefont {Barrett}, \citenamefont {Safronova}, \citenamefont
  {Strouse}, \citenamefont {Tew},\ and\ \citenamefont {Ye}}]{nicholson15}%
  \BibitemOpen
  \bibfield  {author} {\bibinfo {author} {\bibfnamefont {T.}~\bibnamefont
  {Nicholson}}, \bibinfo {author} {\bibfnamefont {S.}~\bibnamefont {Campbell}},
  \bibinfo {author} {\bibfnamefont {R.}~\bibnamefont {Hutson}}, \bibinfo
  {author} {\bibfnamefont {G.}~\bibnamefont {Marti}}, \bibinfo {author}
  {\bibfnamefont {B.}~\bibnamefont {Bloom}}, \bibinfo {author} {\bibfnamefont
  {R.}~\bibnamefont {McNally}}, \bibinfo {author} {\bibfnamefont
  {W.}~\bibnamefont {Zhang}}, \bibinfo {author} {\bibfnamefont
  {M.}~\bibnamefont {Barrett}}, \bibinfo {author} {\bibfnamefont
  {M.}~\bibnamefont {Safronova}}, \bibinfo {author} {\bibfnamefont
  {G.}~\bibnamefont {Strouse}}, \bibinfo {author} {\bibfnamefont
  {W.}~\bibnamefont {Tew}}, \ and\ \bibinfo {author} {\bibfnamefont
  {J.}~\bibnamefont {Ye}},\ }\href {\doibase 10.1038/ncomms7896} {\bibfield
  {journal} {\bibinfo  {journal} {Nature Communications}\ }\textbf {\bibinfo
  {volume} {6}},\ \bibinfo {pages} {6896} (\bibinfo {year} {2015})},\ \Eprint
  {http://arxiv.org/abs/1412.8261} {arXiv:1412.8261} \BibitemShut {NoStop}%
\bibitem [{\citenamefont {Schioppo}\ \emph {et~al.}(2016)\citenamefont
  {Schioppo}, \citenamefont {Brown}, \citenamefont {McGrew}, \citenamefont
  {Hinkley}, \citenamefont {Fasano}, \citenamefont {Beloy}, \citenamefont
  {Yoon}, \citenamefont {Milani}, \citenamefont {Nicolodi}, \citenamefont
  {Sherman}, \citenamefont {Phillips}, \citenamefont {Oates},\ and\
  \citenamefont {Ludlow}}]{schioppo17}%
  \BibitemOpen
  \bibfield  {author} {\bibinfo {author} {\bibfnamefont {M.}~\bibnamefont
  {Schioppo}}, \bibinfo {author} {\bibfnamefont {R.~C.}\ \bibnamefont {Brown}},
  \bibinfo {author} {\bibfnamefont {W.~F.}\ \bibnamefont {McGrew}}, \bibinfo
  {author} {\bibfnamefont {N.}~\bibnamefont {Hinkley}}, \bibinfo {author}
  {\bibfnamefont {R.~J.}\ \bibnamefont {Fasano}}, \bibinfo {author}
  {\bibfnamefont {K.}~\bibnamefont {Beloy}}, \bibinfo {author} {\bibfnamefont
  {T.~H.}\ \bibnamefont {Yoon}}, \bibinfo {author} {\bibfnamefont
  {G.}~\bibnamefont {Milani}}, \bibinfo {author} {\bibfnamefont
  {D.}~\bibnamefont {Nicolodi}}, \bibinfo {author} {\bibfnamefont {J.~A.}\
  \bibnamefont {Sherman}}, \bibinfo {author} {\bibfnamefont {N.~B.}\
  \bibnamefont {Phillips}}, \bibinfo {author} {\bibfnamefont {C.~W.}\
  \bibnamefont {Oates}}, \ and\ \bibinfo {author} {\bibfnamefont {A.~D.}\
  \bibnamefont {Ludlow}},\ }\href {\doibase 10.1038/nphoton.2016.231}
  {\bibfield  {journal} {\bibinfo  {journal} {Nature Photonics}\ }\textbf
  {\bibinfo {volume} {11}},\ \bibinfo {pages} {48} (\bibinfo {year} {2016})},\
  \Eprint {http://arxiv.org/abs/1607.06867} {arXiv:1607.06867} \BibitemShut
  {NoStop}%
\bibitem [{\citenamefont {Tobar}\ \emph {et~al.}(2013)\citenamefont {Tobar},
  \citenamefont {Stanwix}, \citenamefont {McFerran}, \citenamefont {Gu\'{e}na},
  \citenamefont {Abgrall}, \citenamefont {Bize}, \citenamefont {Clairon},
  \citenamefont {Laurent}, \citenamefont {Rosenbusch}, \citenamefont {Rovera},\
  and\ \citenamefont {Santarelli}}]{tobar2013}%
  \BibitemOpen
  \bibfield  {author} {\bibinfo {author} {\bibfnamefont {M.~E.}\ \bibnamefont
  {Tobar}}, \bibinfo {author} {\bibfnamefont {P.~L.}\ \bibnamefont {Stanwix}},
  \bibinfo {author} {\bibfnamefont {J.~J.}\ \bibnamefont {McFerran}}, \bibinfo
  {author} {\bibfnamefont {J.}~\bibnamefont {Gu\'{e}na}}, \bibinfo {author}
  {\bibfnamefont {M.}~\bibnamefont {Abgrall}}, \bibinfo {author} {\bibfnamefont
  {S.}~\bibnamefont {Bize}}, \bibinfo {author} {\bibfnamefont {A.}~\bibnamefont
  {Clairon}}, \bibinfo {author} {\bibfnamefont {P.}~\bibnamefont {Laurent}},
  \bibinfo {author} {\bibfnamefont {P.}~\bibnamefont {Rosenbusch}}, \bibinfo
  {author} {\bibfnamefont {D.}~\bibnamefont {Rovera}}, \ and\ \bibinfo {author}
  {\bibfnamefont {G.}~\bibnamefont {Santarelli}},\ }\href {\doibase
  10.1103/PhysRevD.87.122004} {\bibfield  {journal} {\bibinfo  {journal}
  {Physical Review D - Particles, Fields, Gravitation and Cosmology}\ }\textbf
  {\bibinfo {volume} {87}} (\bibinfo {year} {2013}),\
  10.1103/PhysRevD.87.122004},\ \Eprint {http://arxiv.org/abs/1306.1571}
  {arXiv:1306.1571} \BibitemShut {NoStop}%
\bibitem [{\citenamefont {Dzuba}\ and\ \citenamefont
  {Flambaum}(2017)}]{flambaum17}%
  \BibitemOpen
  \bibfield  {author} {\bibinfo {author} {\bibfnamefont {V.~A.}\ \bibnamefont
  {Dzuba}}\ and\ \bibinfo {author} {\bibfnamefont {V.~V.}\ \bibnamefont
  {Flambaum}},\ }\href {\doibase 10.1103/PhysRevD.95.015019} {\bibfield
  {journal} {\bibinfo  {journal} {Phys. Rev.}\ }\textbf {\bibinfo {volume}
  {D95}},\ \bibinfo {pages} {015019} (\bibinfo {year} {2017})},\ \Eprint
  {http://arxiv.org/abs/1608.06050} {arXiv:1608.06050 [physics.atom-ph]}
  \BibitemShut {NoStop}%
\end{thebibliography}%

\end{document}